\documentclass[12pt]{article}
\usepackage{epsfig,axodraw}
\def\vp{\varphi}
\def\tl{\tilde\lambda}
\def\a{\alpha}
\def\lan{\langle}
\def\ran{\rangle}

\begin{document}
\title{\bf MHV Vertices and Scattering Amplitudes in Gauge Theory}
\author{Jun-Bao Wu \\School of Physics, Peking University \\
Beijing 100871, P. R. China\\ \\
Chuan-Jie Zhu\thanks{Supported in part by fund from the National
Natural Science Foundation of China with grant Number
90103004.} \\
Institute of Theoretical Physics,
Chinese Academy of Sciences\\
P. O. Box 2735,  Beijing 100080, P. R. China}

\maketitle

\begin{abstract}
The generic googly amplitudes in gauge theory are computed by
using the  Cachazo-Svrcek-Witten approach to perturbative
calculation in gauge theory and  the results are in agreement with
the previously well-known ones. Within this approach we also
discuss the parity transformation, charge conjugation and the dual
Ward identity. We also extend this calculation to include fermions
and the googly amplitudes with a single quark-anti-quark pair are
obtained correctly from fermionic MHV vertices. At the end we
briefly discuss the possible extension of this approach to
gravity.
\end{abstract}

\section{Introduction}

Recently Witten \cite{Wittenb} found a deep connection between the
perturbative gauge theory and   string theory in twistor space
\cite{Penrosea}. Based on this work, Cachazo, Svrcek and Witten
reformulated the  perturbative calculation of the scattering
amplitudes in Yang-Mills theory by using the off shell MHV
vertices \cite{Wittena}. The MHV vertices they used are the usual
tree level MHV scattering amplitudes in gauge theory \cite{Parkea,
Giele}, continued off shell in a particular fashion as given in
\cite{Wittena}. (For references on perturbative calculations, see
for example \cite{Parkeb, Berna, Dixon, Bernb, Others}. The 2
dimensional origin of the MHV amplitudes in gauge theory was first
given in \cite{Nair}.) Some sample calculations were done in
\cite{Wittena}, sometimes with the help of symbolic manipulation.
The correctness of the rules was partially verified by reproducing
the known results for small number of gluons \cite{Parkeb}.

In a previous work \cite{Zhu} (for recent works, see
\cite{itp,itpB, Berkovits,BerkovitsMotl,AganagicVafa, Wittenc,
NeitzkeVafa, NekrasovOoguriVafa, GeorgiouKhoze,Gukov, Siegel,
Giombi, Popov,Wittend, Bernc, Kosower, Wittene}), by following the
new approach of \cite{Wittena}, one of the authors computed the
exceptionally simple amplitudes with two positive helicity gluons
and an arbitrary number of negative helicity ones, called googly
amplitudes in \cite{Wittena}. These amplitudes were calculated
from the string theory in \cite{itp}. In the special case when the
two positive helicity particles are adjacent, the result was shown
to be in agreement with the well-known result \cite{Parkea}. In
this paper we will calculate the generic googly amplitudes. By
reproducing the previously well known results, our calculation
gives a quite strong support to the Cachazo-Svrcek-Witten (CSW for
short) proposal.

Although these calculations  gave strong support to the CSW
approach to the perturbative calculation in gauge theory, a direct
proof for the equivalence to the usual Feynman rules seems
hopeless. On the other hand, the twistor   string theory approach
\cite{Wittenb} gives a different but rather compact expression for
the tree level amplitude of gluons \cite{itp,itpB}. It was proved
\cite{itpB} that this expression satisfies most of the
requirements for the tree amplitude of gluons. In \cite{Gukov} it
was argued that the string representation (a connected instanton)
of the tree amplitude can indeed  be decomposed into a summation
of different MHV amplitudes (a set of disconnected instantons),
giving precisely the CSW rules of perturbative calculations.
Encouraged by the success of the googly amplitude calculation we
will present some general discussions on the MHV diagrams in gauge
theory. We note that in the calculation of the googly amplitudes
by using the CSW rules, we only need the MHV vertices with $3$
lines and the MHV vertices with $4$ lines. So in  a parity
conserved theory the higher point (5 or more) MHV amplitudes can
be obtained by parity transformation from the googly amplitudes
and so the higher point MHV vertices can't be chosen arbitrarily.
We also prove that the CSW rules satisfy the dual ward identity
and the charge conjugation identity \cite{Parkeb}.

The CSW proposal \cite{Wittena} can also be extended to gauge
theory with fermions (see also \cite{GeorgiouKhoze}). Although the
MHV (and googly) fermionic amplitudes can be easily obtained by
supersymmetric Ward identities \cite{Grisarua,Grisarub,Parkeb} we
think it is still worthy to compute these amplitudes directly
because the general non-MHV (googly) ammplitudes cannot be
determined in terms of amplitudes only and should be computed
separately\footnote{In an early version of this paper we don't say
this explicitly although we suspect this is the case. See the
footnote of  the recent paper \cite{GeorgiouKhozea}. We thank the
referee to point out this  to us.}. We will compute the googly
amplitudes with fermions by extending the CSW rules with MHV
vertices. For illustration we consider only the simplest case of a
single quark-anti-quark pair. The general cases, including the
supersymmetric case with gluinos, will be discussed in a separate
publication \cite{WuZhu}. Some generic non-MHV fermionic
amplitudes were also computed in
\cite{GeorgiouKhoze,GeorgiouKhozea}.

Another interesting question is whether the CSW approach can be
extended to theories with gravity. A naive extension   of the the
CSW rules to graviton doesn't seem to work \cite{Giombi}.
Nevertheless we believe that some similar rules for graviton must
exist, given the simplicity of the MHV graviton amplitude
\cite{BGK}  and the KLT relations between gauge theory and gravity
\cite{KLT}.  In \cite{Wittend}, Berkovits and Witten put forward a
similar connection between the superconformal supergravity and
closed string theory in twistor space. This may also suggest an
extension of the CSW  rules. In the last section we will present
our partial successful and un-successful attempts to this problem.
In particular we give a simple rule for the calculation of the
off-shell amplitude with a single positive helicity graviton. This
amplitude is proportional to the square of the (only) off-shell
momentum and it is vanishing on shell.

This paper is organized as follows. In section 2, we present the
computation of the generic googly amplitudes. In section 3, we
give some general discussions on the MHV diagrams. We define
precisely how the parity transformation operates in the CSW
approach. We also prove the charge conjugation identity and the
dual Ward identity in this section. In section 4, we extend the
CSW rules to include fermions and compute the googly amplitude
with a single quark and anti-quark pair.  In the last section we
make some investigations on the graviton MHV diagrams. Some
technical proofs are relegated to two Appendices.

\section{The generic googly amplitudes in gauge theory}

First let us recall the rules for calculating tree level gauge
theory amplitudes as proposed in \cite{Wittena}. Here we follow
the presentation given in \cite{Zhu} closely.  We will use the
convention that all momenta are outgoing. By MHV we always mean an
amplitude with precisely two gluons of negative helicity. If the
two gluons of negative helicity are labelled as $r,s$ (which may
be any integers from $1$ to $n$), the MHV vertices (or amplitudes)
are given as follows:
\begin{equation}
V_n =  {\langle\lambda_r,\lambda_s\rangle^4\over
\prod_{i=1}^n\langle\lambda_i, \lambda_{i+1}\rangle} .
\label{eqone}
\end{equation}
For an on shell (massless) gluon, the momentum in bispinor basis
is given as:
\begin{equation}
p_{a\dot a} = \sigma^\mu_{a\dot a} p_\mu = \lambda_a \tilde{
\lambda}_{\dot a}.
\end{equation}
For an off shell momentum, we can no longer define $\lambda_a$ as
above. The off-shell continuation given in \cite{Wittena} is to
choose an arbitrary spinor $\tilde\eta^{\dot a}$ and then to
define $\lambda_a$ as follows:
\begin{equation}
\lambda_a = p_{a\dot a}\tilde{\eta}^{\dot a}.
\end{equation}
For an on shell momentum $p$, we will use the notation
$\lambda_{pa}$ which is proportional to $\lambda_a$:
\begin{equation}
\lambda_{pa} \equiv  p_{a\dot a} \tilde{\eta}^{\dot a} = \lambda_a
\tilde{\lambda}_{\dot a} \tilde{\eta}^{\dot a} \equiv \lambda_a
\phi_p.
\end{equation}
As demonstrated in \cite{Wittena}, it is consistent to use the
same $\tilde\eta$ for all the off shell lines (or momenta). The
final result is independent of $\tilde{\eta}$.

By using only MHV vertices, one can build a tree diagram by
connecting MHV vertices with propagators.  For the propagator of
momentum $p$, we assign a factor $1/p^2$. Any possible diagram
(involving only MHV vertices) will contribute to the amplitude. As
proved in \cite{Wittena}, a tree level amplitude with $n_-$
external gluons of negative helicity must be obtained from an MHV
tree diagram with $n_--1$ vertices. Another relation was given in
\cite{Zhu},
\begin{equation}
n_+=\sum_{i}n_i(i-3)+1,\label{eqnp}
\end{equation}
when $n_+$ is the number of the external gluons with positive
helicity, and $n_i$ is the number of the vertices with exactly $i$
lines. The other relation stated in the above is:
\begin{equation}
n_-=\sum_{i}n_i+1.
\end{equation}
From eq.~(\ref{eqnp}) we can see that  a tree level amplitude with
$n_+$ external gluons of positive helicity will have no
contribution from any diagram containing an MHV vertex with more
than $n_++2$ lines (not necessarily all internal). For the googly
amplitude we have $n_+=2$. Any contributing diagram will have
exactly one MHV vertex with 4 lines. The rest MHV vertices are all
with 3 lines.

In order to compute the googly amplitude we will need the off
shell amplitudes with $n_+=1$. These amplitudes are obtained in
\cite{Zhu}. There are two cases. The first case is when the first
particle with moment $p_1$ is off shell and has positive helicity.
The amplitude is
\begin{equation}
V_n(1+,2-, \cdots, n-) = {p_1^2 \over \phi_2 \phi_n} \, {1 \over
{[}2,3][3,4]\cdots [n-1,n] }. \label{eq34}
\end{equation}
The other case is when the off shell gluon has negative helicity.
We relabel this gluon to be the first one and the amplitude is
given as follows:
\begin{equation}
V_n(1-,2-, \cdots,r+,\cdots,
 n-) = {\phi_r^4 p_1^2 \over \phi_2 \phi_n} \,
{1\over [2,3][3,4]\cdots [n-1,n]} . \label{eq22}
\end{equation}
We stress the fact that the above  off shell amplitudes are
proportional to $p_1^2$ and they  vanish   when $p_1$ is also on
shell ($p_1^2=0$).

\begin{figure}[ht]
    \epsfxsize=100mm%
    \hfill\epsfbox{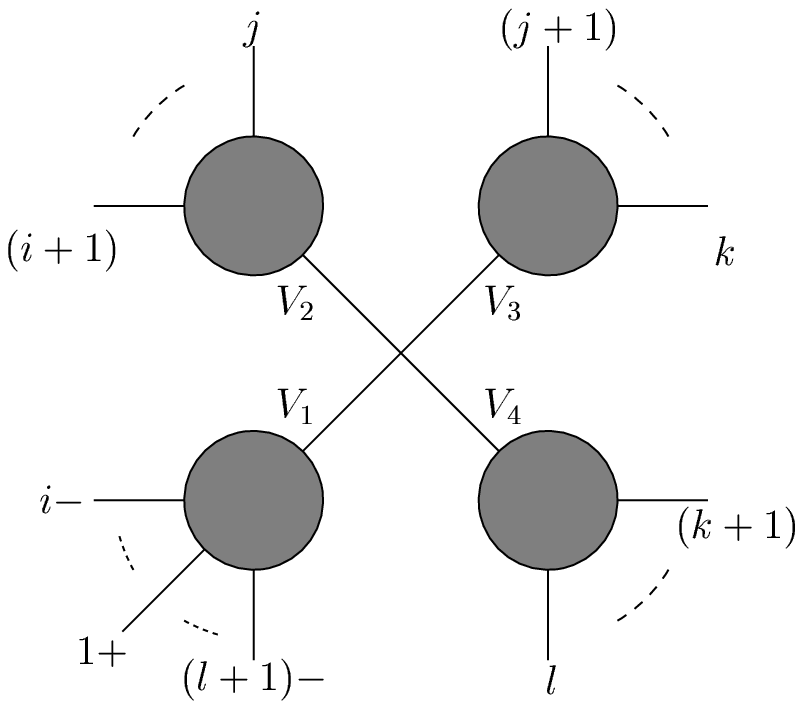}\hfill~\\
    \caption{The diagram decomposition for the generic googly amplitude.
    We
    note that there is only one 4 gluon vertex.}
    \label{figa}
   \end{figure}

Now we compute the generic googly $n$-particle amplitude. The
special case of which when the two positive helicity particles are
adjacent was computed in \cite{Zhu}. The basic idea of the
computation is the same. Now we label the two positive helicity
particles to be $1$ and $r$. As mentioned in the previous section,
it was also proved in \cite{Zhu} that for any googly amplitude the
contributing diagram will have only one 4 line MHV vertex. By
using this result, the amplitude is computed by using the diagram
decomposition as shown in Fig.~\ref{figa}.

\begin{figure}[ht]
    \epsfxsize=100mm%
    \hfill\epsfbox{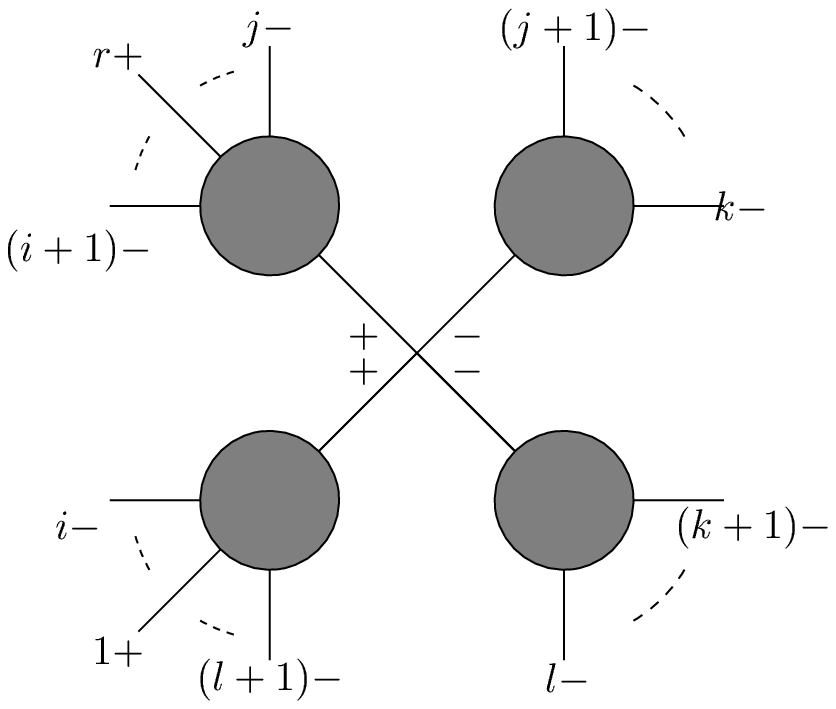}\hfill~\\
    \caption{The diagram decomposition for the generic googly amplitude
    when $(i+1)\le r\le j$.
    }
    \label{figb}
   \end{figure}

\begin{figure}[ht]
    \epsfxsize=100mm%
    \hfill\epsfbox{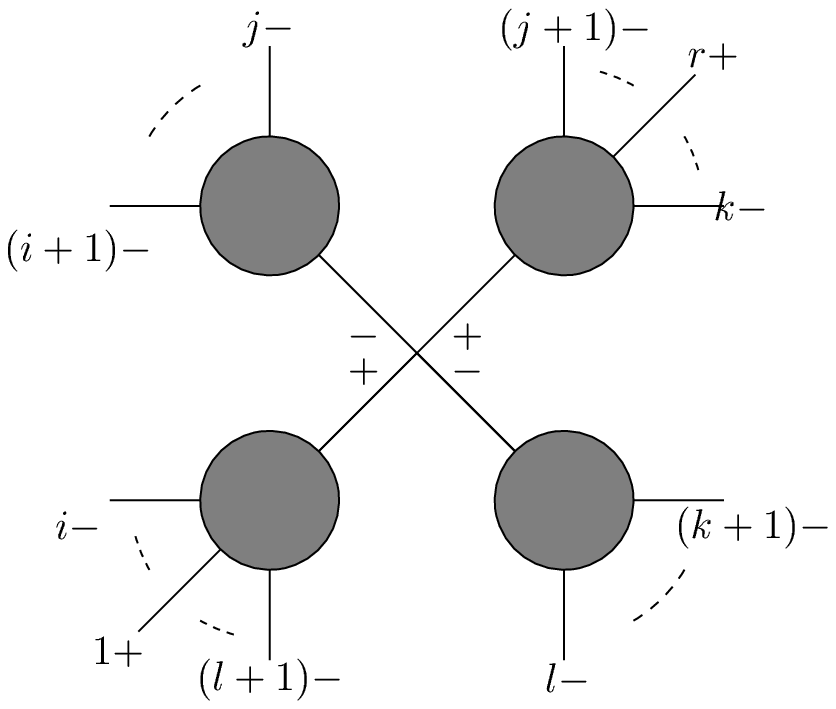}\hfill~\\
    \caption{The diagram decomposition for the generic googly amplitude
    when $(j+1)\le r\le k$.
    }
    \label{figc}
   \end{figure}

\begin{figure}[ht]
    \epsfxsize=100mm%
    \hfill\epsfbox{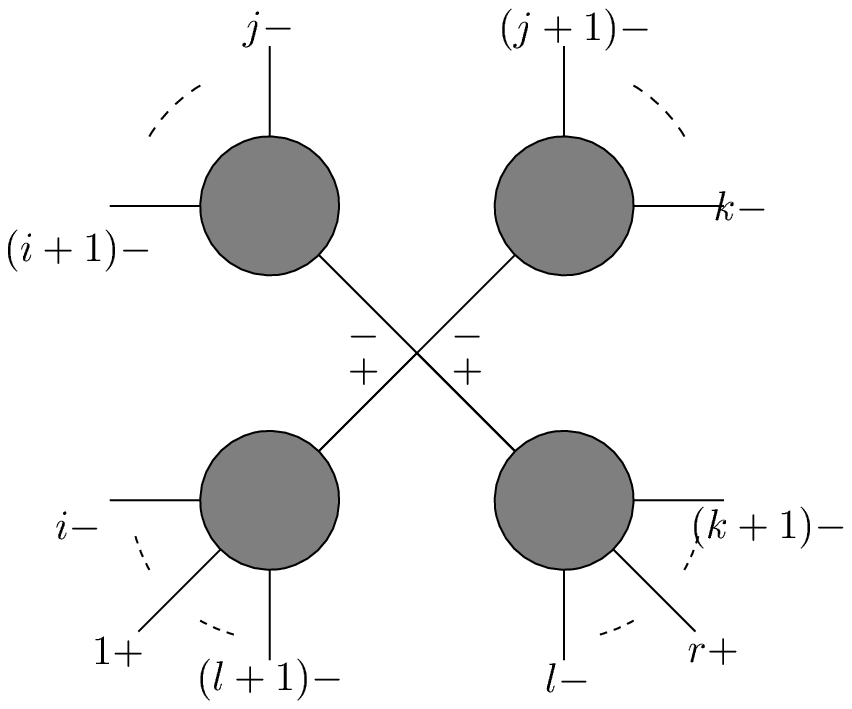}\hfill~\\
    \caption{The diagram decomposition for the generic googly amplitude
    when $(k+1)\le r\le l$.
    }
    \label{figd}
   \end{figure}

For any given $r$, there are three kinds of diagrams as shown in
Fig.~\ref{figb}, Fig.~\ref{figc} and Fig.~\ref{figd}. They depend
on whether the $r$-th gluon is in the second, third  or the last
blob.

All the 4 blob diagrams in these three kinds of diagrams can be
computed by following the method used in \cite{Zhu}. By using the
results in eqs.~(\ref{eq34}) and (\ref{eq22}), the contribution
corresponding to Fig.~\ref{figb} is
\begin{eqnarray}
A_n^1&=&\sum_{i=1}^{r-1}\sum_{j=r}^{n-2}\sum_{k=j+1}^{n-1}\sum_{l=k+1}^n
{\phi_1^4 p_{l+1,i}^2\over  \phi_{l+1} \phi_i}\nonumber \\
&&\times {1 \over [l+1,
l+2]\cdots [n-1, n][n, 1][1,2]\cdots [i-1, i ]}\nonumber \\
&& \times {1\over p_{l+1, i}^2}\times {\phi_r^4 p_{i+1, j}^2\over
\phi_{i+1}\phi_j}{1 \over \prod_{t=i+1}^{j-1}[t, t+1]} \times {1
\over p_{i+1, j}^2} \times {p_{j+1, k}^2\over \phi_{j+1}\phi_k}{1
\over \prod_{t=j+1}^{k-1}[t,t+1]}\nonumber\\
&& \times {1 \over p_{j+1,k}^2}\times {p_{k+1, l}^2\over
\phi_{k+1}\phi_l}\times {1\over \prod_{t=k+1}^{l-1} [t, t+1]} {1
\over p_{k+1, l}^2}\times {\lan V_3, V_4\ran^3\over \lan V_1, V_2
\ran
\lan V_2, V_3\ran \lan V_4, V_1\ran} \nonumber \\
&=&{\phi_1^4 \phi_r^4 \over \prod_{t=1}^n [t, t+1]}
\sum_{i=1}^{r-1}\sum_{j=r}^{n-2}\sum_{k=j+1}^{n-1}\sum_{l=k+1}^n
{[i,i+1]\over \phi_i \phi_{i+1}} \, {[j,j+1]\over \phi_j
\phi_{j+1}} \,\nonumber \\
& & \times{[k,k+1]\over \phi_k \phi_{k+1}} \,{[l,l+1]\over \phi_l
\phi_{l+1}} \,  {\lan V_3, V_4\ran^3\over \lan V_1, V_2 \ran \lan
V_2, V_3\ran \lan V_4, V_1\ran}
\end{eqnarray}
where\footnote{Here and below the summation like
$\sum_{s=l+1}^{n+i}$ is understood as
$\sum_{s=l+1}^n+\sum_{s=1}^i$.}
\begin{eqnarray}
V_1 = \sum_{s=l+1}^{n+i} \lambda_s \phi_s , \qquad & &
V_2 = \sum_{s=i+1}^j \lambda_s \phi_s , \label{eqvii} \\
V_3 = \sum_{s=j+1}^k \lambda_s \phi_s , \qquad & &
 V_4 =\sum_{s=k+1}^l \lambda_s \phi_s ,\label{eqvi}
\end{eqnarray}
and when $i\le j$, we define $p_{i, j}$ as $p_{i, j}=\sum_{t=i}^j
p_t$ as in \cite{Wittena}, when $i>j$, we define $p_{i, j}$ as
$p_{i,j}=\sum_{t=i}^n p_t+\sum_{t=1}^j p_t$.

The contribution corresponding to Fig.~\ref{figc} and
Fig.~\ref{figd} is similar. The result are
\begin{eqnarray}
A_n^2&=&{\phi_1^4 \phi_r^4 \over \prod_{t=1}^n [t, t+1]}
\sum_{i=1}^{r-2}\sum_{j=i+1}^{r-1}\sum_{k=r}^{n-1}\sum_{l=k+1}^n
{[i,i+1]\over \phi_i \phi_{i+1}} \,
{[j,j+1]\over \phi_j \phi_{j+1}} \,\nonumber \\
& & \times{[k,k+1]\over \phi_k \phi_{k+1}} \,{[l,l+1]\over \phi_l
\phi_{l+1}} \,  {\lan V_2, V_4\ran^4 \over \lan V_1, V_2 \ran \lan
V_2, V_3\ran \lan V_3, V_4\ran \lan V_4, V_1\ran}
\end{eqnarray}
and
\begin{eqnarray}
A_n^3&=&{\phi_1^4 \phi_r^4 \over \prod_{t=1}^n [t, t+1]}
\sum_{i=1}^{r-3}\sum_{j=i+1}^{r-2}\sum_{k=j+1}^{r-1}\sum_{l=r}^n
{[i,i+1]\over \phi_i \phi_{i+1}} \,
{[j,j+1]\over \phi_j \phi_{j+1}} \,\nonumber \\
& & \times{[k,k+1]\over \phi_k \phi_{k+1}} \,{[l,l+1]\over \phi_l
\phi_{l+1}} \,  {\lan V_2, V_3\ran^3 \over \lan V_1, V_2 \ran \lan
V_3, V_4\ran \lan V_4, V_1\ran}
\end{eqnarray}
respectively.

The sum of $A_n^i, i=1,2,3$ can be written as
\begin{eqnarray}
A_n & = & {\phi_1^4 \phi_r^4 \over \prod_{t=1}^n [t, t+1]}
\sum_{i=1}^{r-1}\,\sum_{l={\rm max}\{i+3,r\}}^n
 \sum_{j=i+1}^{l-2}\sum_{k=j+1}^{l-1} {[i,i+1]\over \phi_i
\phi_{i+1}} \,
{[j,j+1]\over \phi_j \phi_{j+1}} \,\nonumber \\
& & \times{[k,k+1]\over \phi_k \phi_{k+1}} \,{[l,l+1]\over \phi_l
\phi_{l+1}} \, { \langle V_p, V_q \rangle^4 \over \langle V_1, V_2
\rangle \langle V_2, V_3 \rangle \langle V_3, V_4 \rangle \langle
V_4, V_1 \rangle } , \label{eqab}
\end{eqnarray}
where $p$ and $q$ in eq.~(\ref{eqab}) are the two indexes
($p=2,3,q=3,4,p\ne q$) which satisfy that neither $V_p$ nor $V_q$
includes $\lambda_r\phi_r$ as defined in eqs.~(\ref{eqvii}) and
(\ref{eqvi}).

As in \cite{Zhu}, we can prove that the 4-fold summation in
eq.~(\ref{eqab}) gives exactly the required result, i.e.
\begin{eqnarray}
& & \sum_{i=1}^{r-1}\,\sum_{l={\rm max}\{i+3,r\}}^n
 \sum_{j=i+1}^{l-2}\sum_{k=j+1}^{l-1}  {[i,i+1]\over
\phi_i \phi_{i+1}} \, {[j,j+1]\over \phi_j \phi_{j+1}} \,
{[k,k+1]\over \phi_k \phi_{k+1}} \, {[l,l+1]\over \phi_l \phi_{l+1}} \, \nonumber \\
& & \qquad \qquad \times { \langle V_p, V_q \rangle^4 \over
\langle V_1, V_2 \rangle \langle V_2, V_3 \rangle \langle V_3, V_4
\rangle \langle V_4, V_1 \rangle }=
 {[1,r]^4 \over \phi_1^4 \phi_r^4} . \label{eqac}
 \end{eqnarray}
The proof of this identity, eq.~(\ref{eqac}), will be given in
Appendix A. As in \cite{Zhu}, it is proved by analyzing its pole
terms and finding that all the pole terms are vanishing.

By using eq.~(\ref{eqac}) in eq.~(\ref{eqab}), we have
\begin{equation}
A_n(1+,2-,\cdots, r+,\cdots, n-) = { [1,r]^4 \over \prod_{i=1}^n [
i, i+1] }.
\end{equation}
This is the known result for the googly amplitude \cite{Parkea,
Giele}. It is the complex conjugate of the MHV amplitude,
eq.~(\ref{eqone}), for Minkowski signature.

\section{Some general discussions on the MHV diagrams in gauge theory}

\subsection{Parity transformation and the googly amplitude}

As shown in the previous section, in our calculation of the googly
amplitudes, we use the $3$-line and $4$-line MHV vertices only.
The vertices with more than $4$ lines don't appear in the
contributing diagrams. A parity transformation will exchange the
googly amplitudes with the MHV amplitudes. So in a parity
invariant theory, once the off-shell continuation of $3$ point and
$4$ point MHV amplitudes are given, the on-shell $n$ point MHV
amplitudes ($n>4$) can be obtained from the googly amplitudes by
parity transformation.  The higher MHV  amplitudes (and vertices)
can't be chosen arbitrarily.

The parity invariance of the tree level amplitude is discussed in
\cite{itpB,Wittenc} using  string theory in twistor space
\cite{Wittenb}. Other related works are
\cite{BerkovitsMotl,Berkovits,AganagicVafa,NeitzkeVafa}.

We can make the parity transformation more precise. Denoting the
parity transformation in Minkowski space by ${\cal P}$,  we can
choose its action on the momentum as follows:
\begin{eqnarray}
\lambda_1({\cal P}p)=-\tilde \lambda_2(p), \qquad \lambda_2({\cal
P}p)=\tilde \lambda_1(p).
\end{eqnarray}
This transformation satisfies our basic requirement of changing a
quantity to its complex conjugate:
\begin{equation} \lan
\lambda({\cal P}p_1), \lambda({\cal P}p_2)
\ran=[\tilde\lambda(p_1),\tilde\lambda(p_2)] .
\end{equation}

For the polarization vectors, we have
\begin{equation}
\epsilon_{a\dot a}^+={\mu_a \tilde\lambda_{\dot a}\over \lan \mu,
\lambda \ran}, \qquad \epsilon_{a \dot{a}}^- ={\lambda_a
\tilde\mu_{\dot a}\over [\tilde\lambda, \tilde\mu]} .
\end{equation}
If we choose $\mu_1({\cal P}p)=-\tilde \mu_2(p)$ and $\mu_2({\cal
P}p)=\tilde \mu_1(p)$, then we have $\epsilon^{\mu}({\cal
P}p,h)=P^{\mu}_{\nu} \epsilon^{\nu}(p, -h)$ ($h$ is the helicity).
From this we get
\begin{equation}
A({\cal P}p_1, h_1;\cdots;{\cal P}p_n, h_n)=A(p_1,
-h_1;\cdots;\,p_n,\, -h_n).
\end{equation}
So we have
\begin{eqnarray}
&& \hskip -2cm A(p_1,+;\cdots,p_r,-;\cdots;p_s,-;\cdots;p_n,+)
\nonumber
\\
&&=A({\cal P}p_1,-;\cdots;{\cal P}p_r,+;\cdots;{\cal
P}p_s,+,\cdots,{\cal P}p_n,-)\nonumber \\
&&={[\tilde\lambda({\cal P}p_r),\tilde\lambda({\cal
P}p_s)]^4\over\prod_{i=1}^n[\tilde\lambda({\cal
P}p_i),\tilde\lambda({\cal P}p_{i+1})]}\nonumber\\
&&={\lan \lambda(p_r),\lambda(p_s)\ran^4\over\prod_{i=1}^n\lan
\lambda(p_i),\lambda(p_{i+1})\ran}.
\end{eqnarray}
This shows that a parity transformation indeed exchange the googly
amplitudes with the MHV amplitudes, as we expected.

\subsection{The charge conjugation}

\begin{figure}[ht!]
    \epsfxsize=100mm%
    \hfill\epsfbox{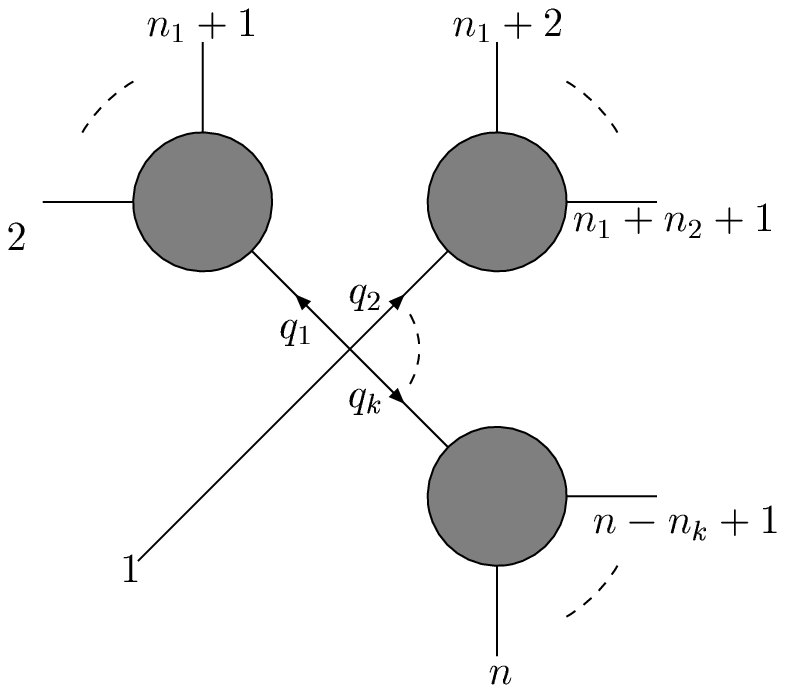}\hfill~\\
    \caption{The diagram decomposition for the gauge amplitude $A(1,\cdots,n)$.}
    \label{figcc2}
   \end{figure}

\begin{figure}[h!]
    \epsfxsize=100mm%
    \hfill\epsfbox{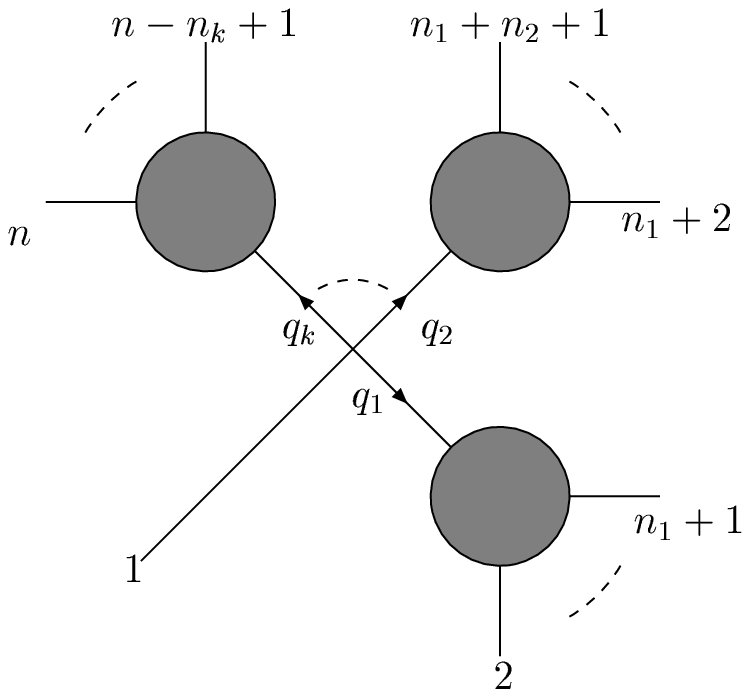}\hfill~\\
    \caption{The diagram decomposition for the gauge amplitude $A(n,\cdots,1)$
     which corresponds
    to the diagram decomposition in Fig.~\ref{figcc2} .}
    \label{figcc1}
   \end{figure}

Charge conjugation can be easily implemented in the CSW approach.
In gauge theory, charge conjugate invariance gives the  following
identity for the tree-level partial amplitudes \cite{Parkeb}:
\begin{equation}
A_n(p_1,h_1;p_2,h_2;\cdots;p_n,h_n)=(-1)^n
A_n(p_n,h_n;\cdots;p_2,h_2;p_1,h_1).\label{eqcc}
\end{equation}
We note that the off shell MHV vertices also satisfies this
identity. Because all the tree level amplitudes are computed from
the MHV vertices  in the framework of CSW, we could easily prove
the above identity by doing charge conjugation operation on the
``Feynman diagrams". We note that Eq.~(\ref{eqcc}) has also been
proved in \cite{itpB} for the case when all particles are on-shell
using the connected instanton  in twistor string theory
\cite{Wittenb}.

We will prove eq.~(\ref{eqcc}) by mathematical induction. The case
for $n=3$ can be easily proven case by case. Assuming  that
eq.~(\ref{eqcc}) is true for all $k<n$ for both on-shell
amplitudes and off-shell amplitudes, we will prove that it is also
true for $k=n$.

The case when the amplitude $A_n(p_1,h_1;p_2,h_2;\cdots;p_n,h_n)$
has less than $2$ gluons with negative helicity are trivial. Since
\begin{equation}A_n(p_1,h_1;p_2,h_2;\cdots;p_n,h_n)=0=(-1)^n
A_n(p_n,h_n;\cdots;p_2,h_2;p_1,h_1).
\end{equation}
When the amplitude $A_n(p_1,h_1;p_2,h_2;\cdots;p_n,h_n)$ is an MHV
amplitude, the amplitude $A_n(p_n,h_n;\cdots;p_2,h_2;p_1,h_1)$ is
an MHV amplitude, too. In this case we can easily prove
eq.~(\ref{eqcc}) by using eq.~(\ref{eqone}).

If the amplitude $A_n(p_1,h_1;p_2,h_2;\cdots;p_n,h_n)$ has more
than $2$ gluons with negative helicity, then the MHV diagrams has
more than $1$ MHV vertices \cite{Wittena}. We can use the diagram
decomposition as in Fig.~\ref{figcc2} to calculate this
amplitude.\footnote{We note that when $n_+=0$, there are no
contributing diagrams and the amplitude vanishes. Then it is
trivial to find that eq.~(\ref{eqcc}) is valid in this case.} The
helicities $h_{q_1},\cdots,h_{q_k}$ of the momenta
$q_1,\cdots,q_k$ must satisfy the constraint that the vertex
$V_{k+1}(p_1,h_1;q_1,h_{q_1};\cdots;q_k,h_{q_k})$ is an MHV
vertex. For every diagram in this diagram decomposition, there is
a unique corresponding diagram in the diagram decomposition used
for calculating $A_n(p_n,h_n;\cdots;p_2,h_2;p_1,h_1)$ as in
Fig.~\ref{figcc1} with the same $h_{q_1},\cdots,h_{q_k}$. In fact
this is a one-to-one correspondence between the diagrams in these
two diagram decompositions.

By using the above diagram composition we have
\begin{eqnarray}
&&\hskip -1.5cm A_n(p_1,h_1;p_2,h_2;\cdots;p_n,h_n)\nonumber\\
&&=\sum V _{k+1}(p_1,h_1;q_1,h_{q_1};\cdots;q_k,h_{q_k})\times
{1\over q_1^2}\times\cdots\times {1\over
q_k^2}\nonumber\\
&&\quad\times A_{n_1+1}(p_2,h_2;\cdots;p_{n_1+1},h_{n_1+1};-q_1,-h_{q_1})\nonumber\\
&&\quad\times\cdots\times
A_{n_k+1}(p_{n-n_k+1},h_{n-n_k+1};\cdots;p_n,h_n;-q_k,-h_{q_k}),\label{eqcc1}
\end{eqnarray}
where
\begin{equation}
q_i=\sum_{j=n_1 + n_2 + \cdots + n_{i-1} + 1}^{n_1 + n_2 + \cdots
+ n_i} p_j ,
\end{equation}
for $1\le i\le k$ and the summation in eq.~(\ref{eqcc1}) is over
$n_1,\cdots,n_k$ subject to the constrains $n_1+\cdots+n_k=n-1$,
$n_i>0$. There is also a summation over all possible helicity
$h_{q_1},\cdots,h_{q_k}$ subject to the constraint that the vertex
$V_{k+1}(p_1,h_1;q_1,h_{q_1};\cdots;q_k,h_{q_k})$ is an MHV
vertex. In the above, the momentum $p_1$ can be off shell.

From the assumed result for all less multi-gluon amplitudes, we
have
\begin{eqnarray}
& &\hskip -1cm A_n(p_1,h_1;p_2,h_2;\cdots;p_n,h_n)=\sum
(-1)^{k+1}V_{k+1}(p_1,h_1;q_k,h_{q_k};\cdots;
q_1,h_{q_1})\nonumber \\
&&\times{1\over q_1^2}\times\cdots\quad\times{1\over
q_k^2}(-1)^{n_1+1}A_{n_1+1}(p_{n_1+1},h_{n_1+1};\cdots;p_2,h_2;-q_1,-h_{q_1})
\nonumber\\
&&\quad\times\cdots
\times(-1)^{n_k+1}A_{n_k+1}(p_n,h_n;\cdots;p_{n-n_k+1},h_{n-n_k+1};-q_k,-h_{q_k})
\nonumber \\
&&=\sum V_{k+1}(p_1,h_1;q_k,h_{q_k};\cdots,q_1,h_{q_1}){1\over
q_1^2}\times\cdots\nonumber \\
&&\quad\times{1\over
q_k^2}A_{n_1+1}(p_{n_1+1},h_{n_1+1};\cdots;p_2,h_2;-q_1,-h_{q_1})\times\cdots\nonumber\\
&&\quad\times
A_{n_k+1}(p_n,h_n;\cdots;p_{n-n_k+1},h_{n-n_k+1};-q_k,-h_{q_k})\nonumber
\\
&&\quad\times (-1)^{k+1+n_1+n_2+\cdots+n_k+k}
\nonumber \\
&&=(-1)^n A_n(p_n,h_n;\cdots;p_2,h_2;p_1,h_1).
\end{eqnarray}
The degenerate case when some $n_i$'s equal to $1$ is also
correctly included in the previous equation. This completes our
proof of eq.~(\ref{eqcc}).

\subsection{The dual Ward identity}

The dual Ward identity is \cite{Giele}
\begin{equation}
\sum_{\sigma\in Z_{n-1}
}A(\sigma(1),\cdots,\sigma(n-1),n)=0.\label{eqdwi}
\end{equation}
where the summation $\sigma$ is over all of the cyclic permutation
of $1,\cdots, n-1$ and the position of $n$ is held fixed to be the
last one by using the cyclic symmetry of the amplitude. This
identity reflect the decoupling of the $U(1)$ degree of freedom
and links the factorization of the partial amplitudes to the
factorization of the full amplitudes \cite{Parkeb}. This has been
discussed in \cite{itpB}. It is also proved in
\cite{ManganoParkeXu} that the dual Ward identity is valid for
(on-shell) MHV amplitudes. We will show that the above dual Ward
identity is valid for all tree-level amplitudes in the framework
of CSW.

Firstly, we   prove that the dual Ward identity is true for the
off shell MHV vertices. By assuming that the two gluons with
negative helicity are $g_p$ and $g_q$, we must prove the
following:
\begin{eqnarray}
& & \hskip -2cm \sum_{\sigma\in Z_{n-1} }
V(\sigma(1),\cdots,\sigma(n-1),n) \nonumber \\
& = & \lan p, q\ran^4 \sum_{\sigma\in Z_{n-1}  } \, {1\over \lan
\sigma(1), \sigma(2)\ran \, \cdots  \lan \sigma(n-1), n \ran \,
\lan n, \sigma(1)\ran} = 0.
\end{eqnarray}
In order to prove this last equality, we note first the following
relation:
\begin{equation} \lan i,
j\ran=\lambda_{i1}\lambda_{j2}-\lambda_{i2}\lambda_{j1}
=\left({\lambda_{i1} \over \lambda_{i2}}-{\lambda_{j1}
\over\lambda_{j2}} \right) \, \lambda_{i2}\lambda_{j2}\, .
\end{equation}
Setting $\psi_i=\lambda_{i1}/\lambda_{i2}$, we have
\begin{eqnarray}
& & \sum_{\sigma\in Z_{n-1}  } \, {1\over \lan \sigma(1),
\sigma(2)\ran \, \cdots  \lan \sigma(n-1), n \ran \, \lan n,
\sigma(1)\ran} = {1\over (\prod_{i=1}^n\lambda_{i2})^2 }  \nonumber \\
& & \hskip 1cm  \times  \, \sum_{\sigma\in Z_{n-1}  } \, {1\over
(\psi_{\sigma(1)} - \psi_{\sigma(2)}) \cdots (\psi_{\sigma(n-1)} -
\psi_n) (\psi_n - \psi_{\sigma(1)}) } .
\end{eqnarray}
So what we need to prove is the following identity:
\begin{equation}
\sum_{\sigma\in Z_{n-1}  } \, {1\over (\psi_{\sigma(1)} -
\psi_{\sigma(2)}) \cdots (\psi_{\sigma(n-1)} - \psi_n) (\psi_n -
\psi_{\sigma(1)}) } = 0, \label{eqdwimhv}
\end{equation}
for arbitrary $\psi_i$'s.

We will prove eq.~(\ref{eqdwimhv}) by mathematical induction. The
case for $n=3$ can be easily checked to be valid. Assuming that it
is valid for $k=n-1$, we will show that it is valid for $k=n$.

We will first prove that the l.h.s. is a constant by showing that
all the pole terms are vanishing.  When $\psi_i\to \infty$ $(1\le
i\le n+1)$, it is evident that the pole terms are vanishing. So we
only need to consider the finite pole terms. The possible finite
pole terms appear when $\psi_j=\psi_n$, $1\le j\le n-1$, or
$\psi_i=\psi_{i+1}$, $1\le i\le n-2$, or $\psi_{n-1}=\psi_1$.

When $\psi_j=\psi_n$, $1\le j\le n-1$, the pole terms in
eq.~(\ref{eqdwimhv}) are from the following two cyclic
permutations:
\begin{equation}
\sigma_{j-1}(1, \cdots, n-1) = (j,j+1, \cdots, n-1, 1,2, \cdots,
j-2,j-1),
\end{equation}
and
\begin{equation}
\sigma_{j}(1, \cdots, n-1) = (j+1,j+2, \cdots, n-1, 1,2, \cdots,
j-1,j) .
\end{equation}
These give 2 pole terms:
\begin{eqnarray}
&& \hskip -1cm {1\over\psi_{j}-\psi_{j+1}}{1
\over\psi_{j+1}-\psi_{j+2}}\cdots{ 1\over \psi_{j-1}-\psi_{n}}
{1\over \psi_{n}-\psi_{j}}+\nonumber\\
&&{1\over \psi_{n}-\psi_{j+1}}{1
\over\psi_{j+1}-\psi_{j+2}}\cdots{ 1\over \psi_{j-1}-\psi_{j}} {
1\over \psi_{j}-\psi_{n}} \nonumber \\
&   =  & {1\over \psi_n - \psi_j} \left[
{1\over\psi_{j}-\psi_{j+1}}{1 \over\psi_{j+1}-\psi_{j+2}}\cdots{
1\over \psi_{j-1}-\psi_{n}} - (\psi_n \leftrightarrow \psi_j)
\right]  .
\end{eqnarray}
The residues in the bracket vanish by taking $\psi_j = \psi_n$.

For $\psi_i= \psi_{i+1}$, $1\le i\le n-2$, the only cyclic
permutation which doesn't give a pole term is:
\begin{equation}
\sigma_i(1,2,\cdots, n-1) = (i+1, i+2, \cdots, n-1, 1, 2, \cdots,
i-1, i). \end{equation} The residues are
\begin{eqnarray}
& & \sum_{ \sigma \ne \sigma_i} {1\over \psi_{\sigma(1)} -
\psi_{\sigma(2)}} {1\over \psi_{\sigma(2)} - \psi_{\sigma(3)}}
\nonumber \\
& & \hskip 1cm \times \cdots {1\over\psi_{i-1}-\psi_{i+1}}
{1\over\psi_{i+1}-\psi_{i+2}}\cdots {1\over \psi_{\sigma(n-1)} -
\psi_n}  {1\over \psi_n  - \psi_{\sigma(1)}} .
\end{eqnarray}
One easily convinces oneself that the above summation is actually
the l.h.s of eq.~(\ref{eqdwimhv}) with the deletion of $\psi_i$
and so it sums to  zero by the assumption of our mathematical
induction. This proves that the residues for $\psi_i = \psi_{i+1}$
vanish. The case for $\psi_{n-1}=\psi_1$ can be proved  by the
same method.

So we proved that all the finite pole terms are vanishing and we
conclude that the l.h.s. must be a constant. This constant can
only be zero because it is homogeneous (with negative degree) in
$\psi_i$'s. This completes our proof of eq.~(\ref{eqdwimhv}).

 \begin{figure}[h!]
    \epsfxsize=100mm%
    \hfill\epsfbox{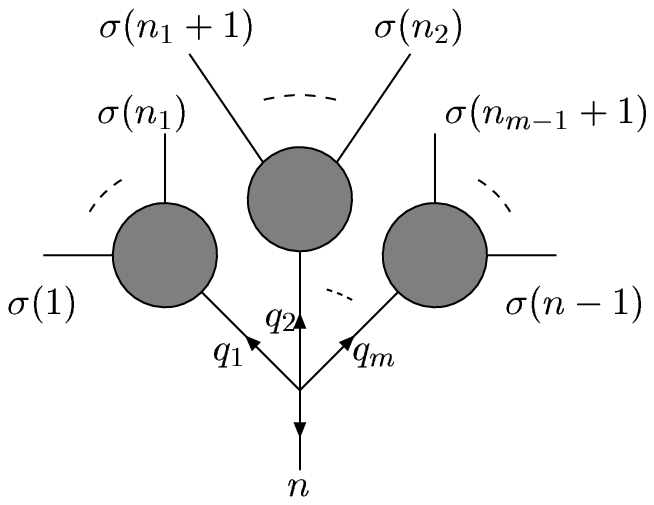}\hfill~\\
    \caption{The decomposition of the amplitude $A(\sigma(1),
    \cdots, \sigma(n-1), n)$ in the proof of the dual Ward
    identity.}
    \label{figdual}
   \end{figure}

By using the above result for dual Ward identity for the off-shell
MHV vertices, one can easily prove the general dual Ward identity,
eq.~(\ref{eqdwi}). The strategy is to write each amplitude
$A(\sigma(1), \cdots, \sigma(n-1), n)$ as a sum over all possible
MHV vertices connected with the $n$-th gluon  as shown in
Fig.~\ref{figdual}. For a fixed MHV vertex, the summation over all
possible cyclic permutations are decomposed as a sum over cyclic
permutations of $q_1$, $q_2$, $\cdots$, $q_m$.\footnote{This
statement should be proved rigorously. We will not try to spell
out the full details of the proof here.} Each sum over cyclic
permutations of $q_1$, $q_2$, $\cdots$, $q_m$ is zero by using the
dual Ward identity for MHV vertex and the general dual Ward
identity then follows.

\section{The googly amplitudes with fermions}

The proposal of \cite{Wittena} can also be extended to gauge
theory with fermions (see also \cite{GeorgiouKhoze}). Although the
MHV (and googly) fermionic amplitudes can be easily obtained by
supersymmetric Ward identities \cite{Grisarua,Grisarub,Parkeb} we
think it is still worthy to compute these amplitudes directly
because the general non-MHV (googly) ammplitudes cannot be
determined in terms of amplitudes only and should be computed
separately.   In this section we will compute the googly
amplitudes with fermions by extending the CSW rules with MHV
vertices. We will consider only  the simplest case of a single
quark-anti-quark pair. The general cases, including the
supersymmetric case with gluinos, will be discussed in a separate
publication \cite{WuZhu}. We also note that some generic non-MHV
fermionic amplitudes were also computed in
\cite{GeorgiouKhoze,GeorgiouKhozea}.

For the case of a single quark-anti-quark pair, the MHV vertices
are as follows (only the $I$-th gluon has negative helicity):
\begin{eqnarray}
& & A(\Lambda_q^+, g_1^+, \cdots, g_I^-, \cdots, g_n^+,
\Lambda_{\bar q}^-) = -{\langle q, I\rangle \langle \bar q, I
\rangle^3 \over \langle q 1\rangle \langle 1,2 \rangle \cdots
\langle n ,\bar q\rangle
\langle \bar q ,q  \rangle} , \\
& & A(\Lambda_q^-, g_1^+, \cdots, g_I^-, \cdots, g_n^+,
\Lambda_{\bar q}^+) = {\langle q, I\rangle^3 \langle \bar q, I
\rangle \over \langle q 1\rangle \langle 1,2 \rangle \cdots
\langle n ,\bar q\rangle \langle \bar q ,q  \rangle} ,
\end{eqnarray}
by denoting the quark with helicity $\pm$ as $\Lambda_q^\pm$ and
the anti-quark as $\Lambda_{\bar q}^\pm$. Gluons are denoted as
$g_i^\pm$ in an obvious notation. What we want to do is to
reproduce the following googly amplitudes (only the $I$-th gluon
has positive helicity):
\begin{eqnarray}
& & A(\Lambda_q^+, g_1^-, \cdots, g_I^+, \cdots, g_n^-,
\Lambda_{\bar q}^-) = {[ q, I]^3 [ \bar q, I ] \over [ q 1] [ 1,2
] \cdots [ n ,\bar q]
[ q ,\bar q  ]} ,  \label{eq45a} \\
& & A(\Lambda_q^-, g_1^-, \cdots, g_I^+, \cdots, g_n^-,
\Lambda_{\bar q}^+) =-{ [ q, I] [\bar q, I ]^3 \over  [ q 1] [ 1,2
] \cdots [ n ,\bar q] [ q ,\bar q  ] }  . \label{eq45b}
 \end{eqnarray}
We use the same off shell continuation as given in \cite{Wittena}
(see also section 2) for off-shell momenta lines. The propagator
for both gluon and gluino internal lines is just $1/p^2$, as
explained in \cite{GeorgiouKhoze}.

We first calculate the amplitudes with a  quark-anti-quark pair
when all gluons have negative helicity and only one particle is
off-shell. By using the same method as in \cite{Zhu}, we   find
that all vertices in the contribution diagrams are 3-line MHV
vertices. When all the $3$ particles are off shell, the MHV
amplitudes with a quark-anti-quark pair can be written as:
\begin{eqnarray}
& &A(\Lambda_q^+, g_1^-, \Lambda_{\bar q}^-) ={\langle1,\bar
q\rangle^2\over\langle q, \bar q \rangle} = \langle q,1 \rangle
=\langle 1,\bar q \rangle =\langle q, \bar q \rangle, \label{eq3f} \\
& &A(\Lambda_q^-, g_1^-, \Lambda_{\bar q}^+)=-{\langle1,\bar
q\rangle^2\over\langle q, \bar q \rangle} =-\langle q,1 \rangle
=-\langle 1,\bar q \rangle =-\langle q, \bar q \rangle
\end{eqnarray}

The amplitudes $A(\Lambda_q^{\pm}, g_1^-, \cdots,
g_n^-,\Lambda_{\bar q}^{\mp})$ when $\Lambda_q^\pm$ is off shell
are given as follows:
\begin{eqnarray}
A(\Lambda_q^+, g_1^-, \cdots, g_n^-,\Lambda_{\bar q}^-)& = &
{p_q^2 \over \phi_1} {1\over [1,2][2,3]\cdots[n-1,n][n,\bar q]},
\label{eqoffshell1} \\
A(\Lambda_q^-, g_1^-, \cdots, g_n^-,\Lambda_{\bar q}^+)& = &
-{p_q^2 \over \phi_1} {1\over [1,2][2,3]\cdots[n-1,n][n,\bar q]},
\end{eqnarray}
and
\begin{eqnarray}
A(\Lambda_q^+, g_1^-, \cdots, g_n^-,\Lambda_{\bar q}^-) & = &
{p_{\bar q}^2 \phi_q^2\over \phi_n}{1\over
[q,1][1,2][2,3]\cdots[n-1,n]},
\label{eqoffshell2}\\
A(\Lambda_q^-, g_1^-, \cdots, g_n^-,\Lambda_{\bar q}^+) & =
 & -{p_{\bar q}^2 \phi_q^2\over \phi_n}{1\over
[q,1][1,2][2,3]\cdots[n-1,n]},
\end{eqnarray}
when the anti-quark $\Lambda_{\bar q}^\pm$ is off-shell. We will
not give the proof of the above formulas here.

When the off shell particle is $g_i$, the result is
\begin{eqnarray}
&&\hskip -2cm A(\Lambda_q^+, g_1^-, \cdots, g_n^-,\Lambda_{\bar
q}^-)={p_i^2\phi_q^3\phi_{\bar q} \over
\phi_{i-1}\phi_{i+1}}\nonumber \\
&&\times{1\over [q,1][1,2]\cdots[i-2,i-1][i+1,i+2]\cdots[n,\bar q]
[\bar q, q]},\label{eqoffshell3}\\
&&\hskip -2cm A(\Lambda_q^-, g_1^-, \cdots, g_n^-,\Lambda_{\bar
q}^+)=-{p_i^2\phi_q^3\phi_{\bar q}\over
\phi_{i-1}\phi_{i+1}}\nonumber\\
&& \times{1\over [q,1][1,2]\cdots[i-2,i-1][i+1,i+2]\cdots[n,\bar
q][\bar q, q]}. \label{eqoffshell4}
\end{eqnarray}
%Here the index $0$ refers to the anti-quark $\Lambda_q$.
Here and in the remaining part of this section the index $0$
refers to $\Lambda_q$ and the index $n+1$ refers to $\Lambda_{\bar
q}$.

 \begin{figure}[h!]
    \epsfxsize=100mm%
    \hfill\epsfbox{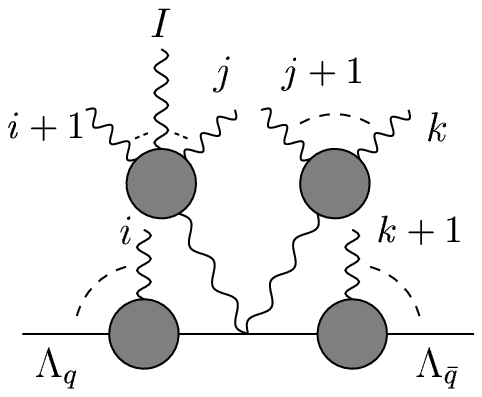}\hfill~\\
    \caption{A sample diagram of 4-line MHV vertex with the quark-anti-quark pair.}
    \label{figcx}
   \end{figure}

Now we begin to compute the googly amplitudes with one
quark-anti-quark pair. Using the same method used in \cite{Zhu},
one can find that in every contributing diagram there is only one
vertex with $4$ lines, all other vertices are with $3$ lines. For
the amplitude  $A(\Lambda_q^+, g_1^-, \cdots, g_I^+, \cdots,
g_n^-, \Lambda_{\bar q}^-)$, one sample diagram was shown in
Fig.~\ref{figcx}. By using the known results for the amplitudes
with exactly one off shell particle given previously in
eqs.~(\ref{eq34})-(\ref{eq22}) and
eqs.~(\ref{eqoffshell1})-(\ref{eqoffshell4}), this gives:
\begin{eqnarray}
 & &\hskip -1cm
 {\phi_q^2 \phi_I^4 \over [q,1][1,2]\cdots [n-1,n][n, \bar
q]}\sum_{i=0}^{I-1}\sum_{j=I}^{n-1}\sum_{k=j+1}^n {[i,i+1]\over
\phi_i \phi_{i+1}} \,{[j,j+1]\over \phi_j \phi_{j+1}}
\,\nonumber \\
& & \times{[k,k+1]\over \phi_k \phi_{k+1}}  \,  {-\lan V_1,
V_3\ran \lan V_4, V_3 \ran^3\over \lan V_1, V_2 \ran \lan V_2,
V_3\ran \lan V_3, V_4\ran \lan V_4, V_1\ran} ,
\end{eqnarray}
where
\begin{eqnarray}
V_1=\sum_{s=0}^i \lambda_s \phi_s, & & \quad V_2=\sum_{s=i+1}^j
\lambda_s \phi_s,\label{eqfv1}\\
 V_3=\sum_{s=j+1}^k \lambda_s \phi_s, & & \quad
V_4=\sum_{s=k+1}^{n+1} \lambda_s \phi_s. \label{eqfv2}
\end{eqnarray}
The complete result is
\begin{eqnarray}
A &=&{\phi_q^2 \phi_I^4 \over [q,1][1,2]\cdots [n-1,n][n, \bar
q]}\sum_{i=0}^{I-1}\,\sum_{k=max\{I, i+2\}}^n\,\sum_{j=i+1}^{k-1}
{[i,i+1]\over \phi_i \phi_{i+1}}
\,\nonumber \\
& & \times {[j,j+1]\over \phi_j \phi_{j+1}}
 \,{[k,k+1]\over \phi_k \phi_{k+1}} \, {-\lan V_1, V_p\ran
\lan V_4, V_p \ran^3\over \lan V_1, V_2 \ran \lan V_2, V_3\ran\lan
V_3, V_4 \ran \lan V_4, V_1\ran} \nonumber \\
& & \hskip -2cm + {\phi_q^3 \phi_{\bar q} \phi_I^4 \over
[q,1][1,2]\cdots [n-1,n][n, \bar q][\bar q,
q]}\sum_{i=0}^{I-1}\,\sum_{l=max\{I,
i+3\}}^n\,\sum_{j=i+1}^{l-2}\sum_{k=j+1}^{l-1} {[i,i+1]\over
\phi_i \phi_{i+1}} \,
\nonumber \\
& &  \hskip -1cm \times {[j,j+1]\over \phi_j \phi_{j+1}}
\,{[k,k+1]\over \phi_k \phi_{k+1}}{[l,l+1]\over \phi_l
\phi_{l+1}}\, {\lan \tilde V_r,\tilde V_s \ran^4\over \lan \tilde
V_1, \tilde V_2 \ran \lan \tilde V_2, \tilde V_3\ran \lan \tilde
V_3,\tilde V_4 \ran \lan \tilde V_4,\tilde V_1\ran},\label{eqfg2}
\end{eqnarray}
where
\begin{eqnarray}
&& \tilde V_1=\sum_{s=0}^i \lambda_s \phi_s+\sum_{s=l+1}^{n+1}
\lambda_s \phi_s,  \quad \tilde V_2=\sum_{s=i+1}^j \lambda_s
\phi_s, \label{eqftv1}\\
&& \tilde V_3=\sum_{s=j+1}^k \lambda_s \phi_s,  \quad \tilde
V_4=\sum_{s=k+1}^l \lambda_s \phi_s . \label{eqftv2}
\end{eqnarray}
The $p$ in eq.~(\ref{eqfg2}) is the index ($p=2,3$) which
satisfies that $V_p$ doesn't include $\lambda_I \phi_I$ as defined
in eqs.~(\ref{eqfv1}) and (\ref{eqfv2}) and the $r$ and $s$ in
eq.~(\ref{eqfg2}) are the two indices ($r=2,3,s=3,4,r\ne s$) which
satisfy that neither $\tilde V_r$ nor $\tilde V_s$ includes
$\lambda_I\phi_I$ as defined in eqs.~(\ref{eqftv1}) and
(\ref{eqftv2}). The summations in eq.~(\ref{eqfg2}) can be done by
using identities involving $\lambda_i$ as in the gluon googly
amplitude. After doing this summation we obtained the correct
googly amplitude with a single quark-anti-quark pair as given in
eqs.~(\ref{eq45a}) and (\ref{eq45b}). The details for the relevant
identities and also the extension to multi-pairs of
quark-anti-quark pairs and gluinos will be given in a separate
publication \cite{WuZhu}.

\section{The MHV diagrams for gravity}

Encouraged by the success of the CSW approach to gauge theory, it
is natural to ask  if a similar approach to gravity exists. We
expected so also because there exist some close relations between
the gravity amplitudes and the gauge amplitudes, the famous KLT
relations \cite{KLT}.

We use the same off-shell continuation
$\lambda_a=p_{a\dot{a}}\tilde \eta^{\dot{a}}$ as in
\cite{Wittena}. The off shell MHV vertex with 3 gravitons is
\cite{BGK}
\begin{equation}
V_3^{graviton}(1+,2-,3-)=\left({\langle 2,3\rangle^3 \over \langle
1,2 \rangle \langle 3,1 \rangle}\right)^2,
\end{equation}
and the propagator for the graviton with momentum $p$ is still
$1/p^2$.

As in gauge theory, we first compute $A(1+,2-,\cdots,n-)$ when
only one particle is off-shell. We will give a formula for
$A(1+,2-,\cdots,n-)$ when only the momentum $p_i$ is off-shell.
This amplitude vanishes when all momenta are on shell.  So we
expect that it is proportional to $p_i^2$.  This turns out to be
true for gravity. We note that the vanishing for the $4$-particle
and $5$-particle on-shell amplitudes were also obtained in
\cite{Giombi}.

The diagrams which contribute to $A(1+,2-,\cdots,n-)$   include
the MHV vertices with $3$ lines only. By using momentum
conservation, one can show that the  MHV  vertex $V_3(1+,2-,3-)$
is actually polynomial in $\lambda$:
\begin{equation}
V_3(1+,2-,3-)=\langle \lambda_1,\lambda_2\rangle^2=\langle
\lambda_2,\lambda_3\rangle^2=\langle \lambda_3,\lambda_1\rangle^2.
\label{eq3gravitons}
\end{equation}
In order to present the rules for the computation of $A(1+,2-,
\cdots, n-)$, we will need   some basic concepts and notations
from  graph theory.  By a graph $\Gamma$, we mean two set: the
vertex set $V$ and the edge set $E\subseteq\{e_{ij} | i,j\in V,
i\ne j\}$. We will use the undirected graph only. This means that
$e_{ij}=e_{ji}$ for all $i,j\in V$. We will denote the vertex set
of graph $\Gamma$ as $V(\Gamma)$ and denote the edge set of graph
$\Gamma$ as $E(\Gamma)$.  A connected graph not containing any
cycles is called a tree. Two vertex $i,j$ are adjacent if there is
a edge $e_{ij}$ in $E(\Gamma)$.

We relabel the off-shell graviton to be the first one. The formula
for the amplitudes is:
\begin{equation}
A(p_1,h_1;p_2,h_2;\cdots;p_n,h_n)=p_1^2
\sum_{\Gamma,V(\Gamma)=S(n)} P(\Gamma),\label{eqgamma}
\end{equation}
where  $h_i=\pm 2$ is the helicity of the graviton whose momentum
is $p_i$ and $\Gamma$ is undirected tree whose vertex set is
$V(\Gamma)=S(n)\equiv\{2,3,\cdots,n\}$ and $P(\Gamma)$ can be
obtained from the following rules:
\begin{itemize}

\item  For every vertex $i$ in $V(\Gamma)$, there is a factor
$p(i)=\phi_i^{2 h_i}$. It means that if the graviton with momentum
$p_i$ has positive helicity, the factor is $\phi_i^4$; otherwise
the factor is $\phi_i^{-4}$. We stress the fact that  there is
only one positive helicity graviton.

\item  For every edge $e_{ij}$ in $E(\Gamma)$ there is a factor
$p(e_{ij})$:
\begin{equation}
p(e_{ij})=\phi_i^2\, \phi_j^2\,
 {\langle i,j\rangle \over [i,j]} .
\end{equation}
\end{itemize}

\noindent  $P(\Gamma)$ is the product of above factors:
\begin{equation}
P(\Gamma)=\prod_{i\in V(\Gamma)} p(i) \prod_{e_{ij}\in
E(\Gamma)}p(e_{ij}).
\end{equation}
By using the above rules we have
\begin{equation}
A(p_1,h_1,\cdots,p_n,h_n)=p_1^2 \prod_{i=2}^n\phi_i^{2
h_i}\sum_{\Gamma,V(\Gamma)=S(n)} \prod_{e_{ij}\in
E(\Gamma)}p(e_{ij}). \label{eq333}
\end{equation}
The summation in eq.~(\ref{eq333}) is over all of the inequivalent
undirected trees with vertex set $S_n$.  The result can be proved
by mathematical induction. The detail is relegated to  Appendix B.

The on-shell 4-graviton MHV amplitude is \cite{BGK}:
\begin{equation}
A_4(1+,2+,3-,4-)={\langle 3,4\rangle^8\over \langle 1,2 \rangle
\langle 1,3 \rangle\langle 1,4 \rangle\langle 2,3 \rangle \langle
2,4 \rangle \langle 3,4 \rangle}{[3,4]\over \langle
1,2\rangle}.\label{eq4gravitons}
\end{equation}
We note that a straightforward off-shell continuation
$\lambda_a=p_{a\dot{a}}\tilde\eta^{\dot{a}}$ is not possible for
$A_4$ because of the existence of the ``anti-holomorphic term"
$[3,4]$. By using the on-shell condition, like $p_3^2=p_4^2=0$,
and momentum conservation,  one can write this term in various
forms which are equivalent on shell and include the ``holomorphic
terms" and the momenta product  only.

For example, we can use $\langle 3,4\rangle[3,4]=2 p_3\cdot p_4$
to write $[3,4]$ as $2 p_3\cdot p_4/\langle 3,4\rangle$. On the
other  hand, using the on-shell condition $p_i^2=0,i=1,\cdots,4$
and the momentum conservation, we have $2 p_3\cdot p_4=p_{34}^2=2
p_1\cdot  p_2$. We can use  these relations to give    two more
way of writing the ``non-holomorphic" expression $[3,4]$.

We can also use $\langle 1,2\rangle[2,3]+\langle1,4\rangle
[4,3]=0$ to write
\begin{equation}
{[3,4]\over\langle 1,2\rangle}={[2,3]\over\langle
1,4\rangle}={p_{23}^2\over \langle 2,3\rangle \langle 1,4\rangle},
\end{equation}
or use $\langle 1,3\rangle[3,4]+\langle1,2\rangle [2,4]=0$ to
write
\begin{equation}
{[3,4]\over\langle 1,2\rangle}={[4,2]\over\langle
1,3\rangle}={p_{13}^2\over \langle 1,3\rangle \langle 4,2\rangle}.
\end{equation}
One can try to write all the possible (on shell) equivalent forms
for  the  4-graviton MHV amplitude. There are about 9 different
forms  which satisfy the obvious symmetry. Nevertheless we have
not been able to obtain the correct 5-graviton amplitude by using
any one of these forms or one of their linear combination. It is
safe to conclude that a naive extension of the CSW approach to
gravity failed (see also \cite{Giombi}). New ingredient must be
introduced in order to develop a CSW like rules for gravity. The
similar connection between conformal supergravity and twistor
string theory discussed in the recent paper \cite{Wittend}  may
offer some clues.

To extend the CSW approach further,  we may change the 3-particle
vertex. A possible  ``MHV vertex" for charged particle is
\begin{eqnarray}
A(1+,2-,3-)&  = & \left({\langle 2,3\rangle^3 \over \langle 1,2
\rangle \langle 3,1 \rangle}\right)^{2s-1} \nonumber \\
& = & \langle 1,2\rangle^{2s-1}=\langle 2,3\rangle^{2s-1}=\langle
3,1\rangle^{2s-1} ,
\end{eqnarray}
where $s$ is a positive integer. By using this vertex and the CSW
rules, one can also compute $A(1+,2-,3-,4-)$ and the result is
\begin{eqnarray} A(1+,2-,3-,4-)&=&  {(\phi_2\phi_3\phi_4)}^{-(2s-1)}  \times
\left( {\langle 2,3 \rangle^{2s-2}\over [2,3]}\langle
\lambda_{23},
\lambda_{p4} \rangle^{2s-1} \right. \nonumber\\
  & & \left. + {\langle 3,4
\rangle^{2s-2}\over [3,4]}\langle \lambda_{p2},  \lambda_{34}
\rangle^{2s-1}\right) .
\end{eqnarray}
By using the following equation:
\begin{equation}
{\langle \lambda_{23}, \lambda_{p4} \rangle \over [2,3]}+{\langle
\lambda_{p2}, \lambda_{34} \rangle\over [3,4]}=0,
\end{equation}
we have
\begin{eqnarray} A(1+,2-,3-,4-)&=&{1\over
[2,3]^{2s-1}}\left[(p_2\cdot p_3)^{2s-2}-(p_3\cdot
p_4)^{2s-2}\right]
\nonumber\\
& & \times  \langle \lambda_{23}, \lambda_{p4}
\rangle^{2s-1}{(\phi_2\phi_3\phi_4)}^{-(2s-1)} .
\end{eqnarray}
So when $s>1$ this amplitudes doesn't vanish for generic
$\lambda_i,\tilde \lambda_i(1\le i \le 4)$. This indicates that
the CSW approach can't be arbitrarily extended to include higher
derivative theories. It remains to discover the hidden principle
behind the success of the CSW approach to gauge theory and its
coupling to fermions, at least for tree level calculations.

\section*{Acknowledgments}

We would like to thank Zhe Chang, Bin Chen, Han-Ying Guo, Miao Li,
Jian-Xin Lu, Jian-Ping Ma, Ke Wu and Yong-Shi Wu  for discussions.
Jun-Bao Wu would like to thank Qiang Li for help on drawing the
figures. Chuan-Jie Zhu would like to thank Jian-Xin Lu  and the
hospitality at the Interdisciplinary Center for Theoretical Study,
University of Science and Technology of China where part of this
work was done.

\section*{Appendix A: The proof of eq.~(\ref{eqac})}

In this appendix we will prove eq.~(\ref{eqac}). As in \cite{Zhu},
we can use a $SL(2,{\bf C})$ transformation and a rescaling of
$\tilde\eta$ to choose $\tilde\eta^1=0$ and $\tilde\eta^2=1$. Then
we have
\begin{equation}
 \phi_i=\tl_{i2}
\end{equation}
and
\begin{equation}
{[i,j]\over
\phi_i\phi_j}={{\tl_{i1}\tl_{j2}-\tl_{i2}\tl_{j1}}\over
\tl_{i2}\tl_{j2}}={\tl_{i1}\over\tl_{i2}}-{\tl_{j1}\over\tl_{j2}}.
\end{equation}

If we do a rescaling of $\tilde\lambda_{i1}$ by
$\tilde\lambda_{i2}$, i.e. by defining $\vp_i =
{\tilde\lambda_{i1} \over \tilde\lambda_{i2}} $, and also do a
rescaling of $\lambda_{ia}$ by $1/\tilde\lambda_{i2}$, then
eq.~(\ref{eqac}) becomes:
\begin{eqnarray}
F(\vp)&=&\sum_{i=1}^{r-1}\,\sum_{l={\rm max}\{i+3,r\}}^n
 \sum_{j=i+1}^{l-2}\sum_{k=j+1}^{l-1}
(\vp_i-\vp_{i+1})(\vp_j-\vp_{j+1})\nonumber
\\
&&   \times  (\vp_k-\vp_{k+1})(\vp_l-\vp_{l+1}) {\langle
V_p,V_q\rangle^4\over\langle V_1, V_2\rangle\langle V_2,
V_3\rangle\langle V_3, V_4\rangle\langle V_4, V_1\rangle}
\nonumber \\
&  = & (\vp_1-\vp_r)^4. \label{eq17}
\end{eqnarray}
where
\begin{equation}
V_1 = \sum_{s=l+1}^{n+i} \lambda_s, \qquad V_2= \sum_{s=i+1}^j
\lambda_s,\label{eqv12}
\end{equation}
\begin{equation}
V_3 = \sum_{s=j+1}^k \lambda_s, \qquad V_4= \sum_{s=k+1}^l
\lambda_s. \label{eqv34}
\end{equation}
There are also two constraints:
\begin{eqnarray}
 & & V_1 + V_2 + V_3 + V_4 = \sum_{l=1}^n \lambda_l = 0 , \label{eq29} \\
 & & \sum_{l= 1}^n \lambda_i\,  \varphi_i = 0 , \label{eq30}
 \end{eqnarray}
from momentum conservation.

From eq.~(\ref{eq29}) and eq.~(\ref{eq30}) we can solve
$\lambda_1$ and $\lambda_r$ in terms of the rest $\lambda_{i}$'s
and all $\varphi_j$'s  as:
\begin{eqnarray}
&& \lambda_1=-\sum_{2\le j\le n, j \ne
r}{\vp_j-\vp_r\over\vp_1-\vp_r}\lambda_j,\label{eqla1}\\
&&\lambda_r=\sum_{2\le j\le n, j \ne
r}{\vp_j-\vp_1\over\vp_1-\vp_r}\lambda_j.\label{eqla2}
\end{eqnarray}
By using the above result, the left hand side of eq.~(\ref{eq17})
can be considered as a function of $\lambda_{j}$ ($2\le j\le n$,
$j\ne r$) and all $\varphi_j$. As a function of $\varphi$ we will
show that it is independent of $\varphi_j$ for $2\le j\le n$ and
$j\ne r$.

First we  show  that there is no pole terms for $\varphi_s \to
\infty, 2\le s\le n-1, s \ne r$. We will show that every term in
the l.h.s of eq.~(\ref{eq17}) will not tend to infinity when
$\varphi_s \to \infty$. We denote
$(\varphi_i-\varphi_{i+1})(\varphi_j-\varphi_{j+1})(\varphi_k-\varphi_{k+1})
(\varphi_l-\varphi_{l+1})$  in eq.~(\ref{eq17}) as $F(i,j,k,l)$
and $\langle V_p,V_q\rangle^4/( \langle V_1, V_2\rangle\langle
V_2, V_3\rangle\langle V_3, V_4\rangle\langle V_4, V_1\rangle)$ as
$A_4(V_1,V_2,V_3,V_4)$.

When $\varphi_s \to \infty$, $\lambda_1$ and $\lambda_r$ will grow
as $\varphi_s$ and other $\lambda_i$'s does not depend on
$\varphi_s$. $V_1$ and the $V_t$ which includes $\lambda_r$ as
defined in eqs.~(\ref{eqv12}) and (\ref{eqv34}) will tend to
infinity as fast as $\varphi_s$. $V_p$ and $V_q$ are independent
of $\varphi_s$. So the numerator of $A_4(V_1,V_2,V_3,V_4)$ is
independent of $\varphi_s$.

Now we prove  that $\langle V_1,V_t \rangle$ grows as $\varphi_s$.
From eq.~(\ref{eqla1}) and eq.~(\ref{eqla2}), we can write
$\lambda_1$ and $\lambda_r$ as $\lambda_1=a_{1s}
\varphi_s\lambda_s+\mu_{1s}$ and $\lambda_r=a_{rs} \varphi_s
\lambda_s+\mu_{rs}$ respectively, where $\mu_{1s}$ and $\mu_{rs}$
is independent on $\varphi_s$. Then $\langle V_1,V_t
\rangle=\varphi_s\langle \lambda_s, a_{1s} \mu_{rs}-a_{rs}
\mu_{1s} \rangle+O(\varphi_s^0)$ will grow as $\varphi_s$ for
generic $\varphi_i,\lambda_j, 1\le j\le n, j\ne r$.

The factor $F(i,j,k,l)$ grows as $\varphi_s^2$ at most. Also  one
can check case by case and find the denominator of
$A_4(V_1,V_2,V_3,V_4)$  grows as $\varphi_s^2$ at
least\footnote{We note that when one of $V_p$ and $V_q$ includes
only the term $\lambda_s$, the denominator will grows as
$\varphi_s^2$ because $\langle V_1,\lambda_s\rangle$ and $\langle
V_t,\lambda_s\rangle$ are independent of $\varphi_s$.}.  By
combing all these results one sees that  there is no pole terms
for $F(\varphi)$ as $\varphi_s \to \infty$.

The next step is to show that all the finite pole terms in
$F(\varphi)$ are vanishing. The possible terms appear if any
factor of $\langle V_1, V_2\rangle\langle V_2, V_3\rangle\langle
V_3, V_4\rangle\langle V_4, V_1\rangle$ is vanishing.

Let us consider first the vanishing of $\langle V_1,V_2\rangle$.
We denote this set of $V_1$ and $V_2$ as $v_1$ and $v_2$:
\begin{equation}
v_1=\sum_{i=n_3+1}^{n+n_1}\lambda_i, \qquad
v_2=\sum_{i=n_1+1}^{n_2}\lambda_i.
\end{equation}
First let us consider the case when $n_1+1\le r\le n_2$.

\begin{figure}[h!]
    \epsfxsize=100mm%
    \hfill\epsfbox{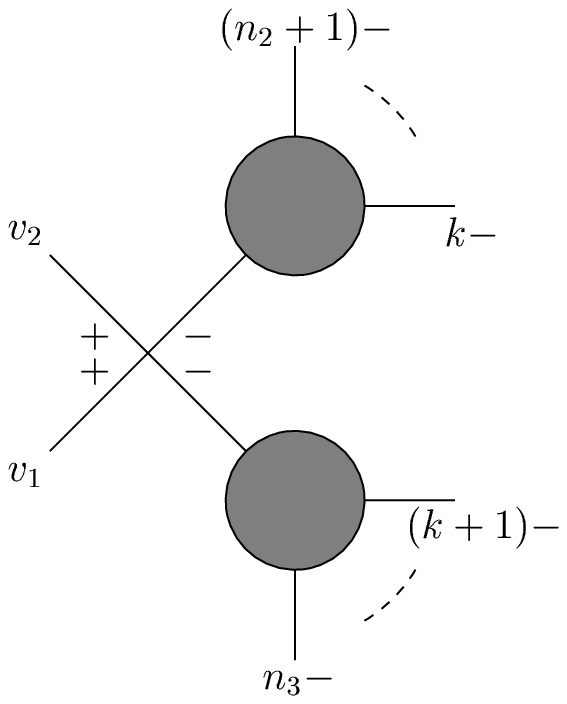}\hfill~\\
    \caption{The pole terms $\langle v_1,v_2\rangle$ from the factor $\langle V_1,
    V_2\rangle$. There is a summation over $k$.   }
    \label{figcc}
   \end{figure}

As one can see from Fig.~\ref{figcc} that there is contribution
from summing over $k$ and fixing $i=n_1$, $j=n_2$ and
$l=n_3$.\footnote{We note that $V_3$ and $V_4$ depend on $k$:
$V_3=\sum_{i=n_2+1}^{k}\lambda_i$ and $V_4 = \sum_{i=k+1}^{n_3}
\lambda_i $.} The residues (ignoring an overall factor
$(\vp_{n_1}-\vp_{n_1+1})(\vp_{n_2}-\vp_{n_2+1})(\vp_{n_3}-\vp_{n_3+1})$)
for this pole terms are
\begin{equation}
C_1=\sum_{k=n_2+1}^{n_3-1}(\vp_k-\vp_{k+1}){\langle V_3, V_4
\rangle^3 \over \langle v_2, V_3 \rangle \langle V_4, v_1\rangle}.
\end{equation}

Because $\langle v_1, v_2 \rangle=0$, $v_1$ and $v_2$ are linearly
dependent, we can assume that $v_i=\a_i v_0, i=1,2$, for some
$\a_i$ and $v_0$.

Using this result and $v_1+v_2+V_3+V_4=0$, we have
\begin{eqnarray}
C_1={(\a_1+\a_2)^3 \over
\a_1\a_2}\sum_{k=n_2+1}^{n_3-1}(\vp_k-\vp_{k+1})\langle v_0, V_3
\rangle. \label{eqc1}
\end{eqnarray}
By using a little algebra, we can write the summation in
eq.~(\ref{eqc1}) as following
\begin{eqnarray}
C_1={(\a_1+\a_2)^3 \over \a_1\a_2}
\left(\sum_{k=n_2+1}^{n_3-1}\vp_k\langle v_0, \lambda_k \rangle
-\vp_{n_3}\langle v_0, \sum_{m=n_2+1}^{n_3-1}\lambda_m\rangle
\right).
\end{eqnarray}
Because $\lan v_0, \sum_{m=n_2+1}^{n_3}\lambda_m\ran=\lan v_0$,
$V_2\ran=0$, we have $\lan v_0,
\sum_{m=n_2+1}^{n_3-1}\lambda_m\ran=-\lan v_0, \lambda_{n_3}\ran$.
Then
\begin{equation}
C_1={(\a_1+\a_2)^3\over \a_1\a_2}\lan v_0, \sum_{k=n_2+1}^{n_3}
\vp_k \lambda_k\ran.
\end{equation}

   \begin{figure}[h!]
    \epsfxsize=100mm%
    \hfill\epsfbox{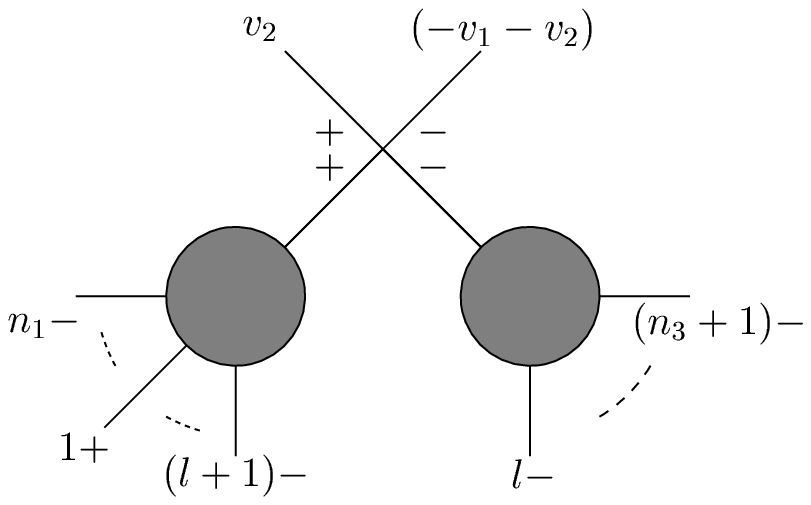}\hfill~\\
    \caption{The pole terms $\langle v_1,v_2\rangle$ from the factor $\langle V_2,
    V_3\rangle$. There is a summation over $l$.   }
    \label{figcd}
   \end{figure}

Unlike  the special case discussed in \cite{Zhu}, here for
$n_1+1\le r\le n_2$, there are pole terms from the vanishing of
the factor $\lan V_2, V_3 \ran$ in Fig.~\ref{figcd} by setting
$V_2=v_2$ and $V_3=-v_1-v_2$ and summing over $l$. This gives the
following contribution:
\begin{eqnarray}
C_2&=&\sum_{l=n_3+1}^n(\vp_l-\vp_{l+1}){\lan -v_1-v_2 ,
V_4\ran^3\over \lan v_1-V_4, v_2\ran \lan V_4, v_1-V_4 \ran}
\nonumber \\ &=&   {(\a_1+\a_2)^3\over
\a_1\a_2}\sum_{l=n_3+1}^n(\vp_l-\vp_{l+1})\lan v_0, V_4\ran
\nonumber \\
&=&  {(\a_1+\a_2)^3\over \a_1\a_2}\lan v_0, \sum_{l=n_3+1}^n \vp_l
\lambda_l-\vp_1 \sum_{m=n_3+1}^n \lambda_m\ran.
\end{eqnarray}

 \begin{figure}[h!]
    \epsfxsize=100mm%
    \hfill\epsfbox{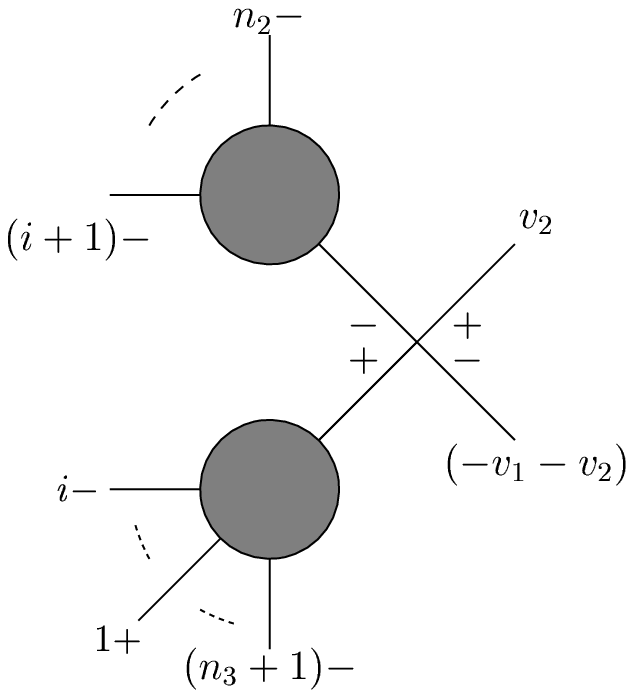}\hfill~\\
    \caption{The pole terms $\langle v_1,v_2\rangle$ from the factor $\langle V_3,
    V_4\rangle$. There is a summation over $i$.   }
    \label{figce}
   \end{figure}
The third piece of the pole terms is from the vanishing of $\lan
V_3, V_4 \ran$ in Fig.~\ref{figce} by setting $V_3=v_2$ and
$V_4=-v_1-v_2$ and summing over $i$. This give the following
contribution:
\begin{eqnarray}
C_3={(\a_1+\a_2)^3\over \a_1\a_2} \lan
v_0,\vp_1\sum_{m=n_3+1}^{n+1}{\lambda_m}+\sum_{i=2}^{n_1}\vp_i\lambda_i\ran
\end{eqnarray}

\begin{figure}[h!]
    \epsfxsize=100mm%
    \hfill\epsfbox{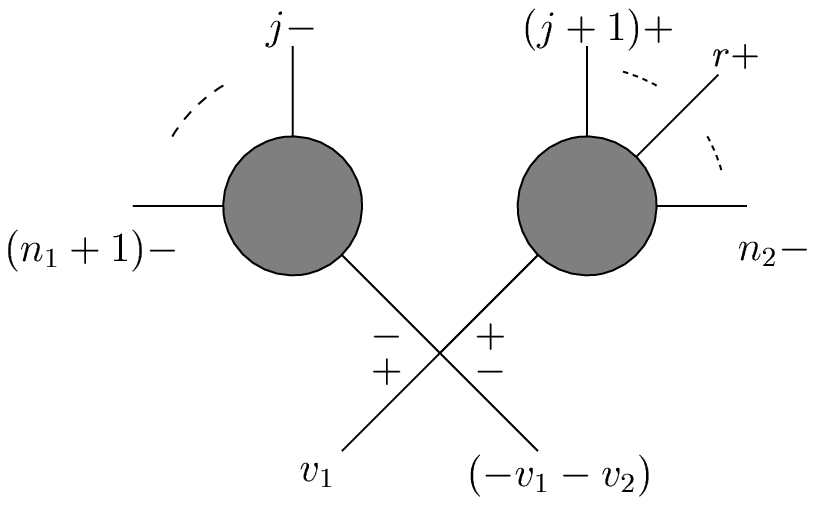}\hfill~\\
    \caption{The pole terms $\langle v_1,v_2\rangle$ from the factor $\langle V_4,
    V_1\rangle$. There is a summation over $j$ from $n_1+1$ to $r-1$.}
    \label{figcf}
   \end{figure}

\begin{figure}[h!]
    \epsfxsize=100mm%
    \hfill\epsfbox{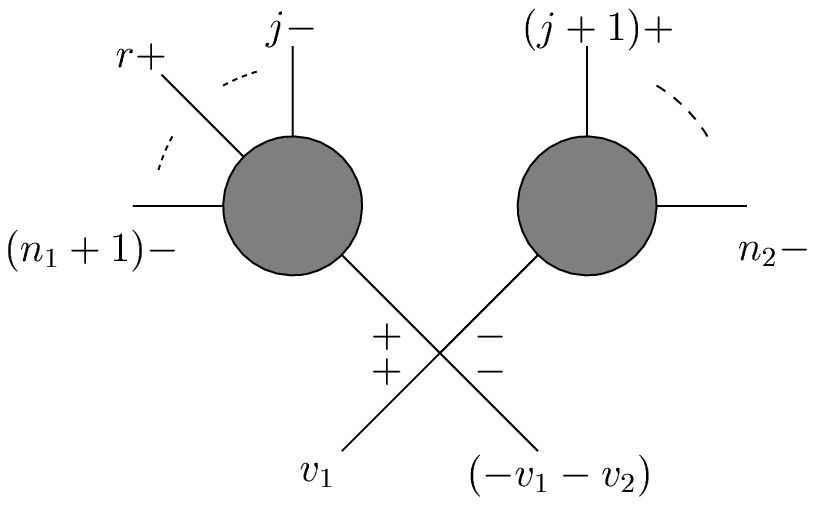}\hfill~\\
    \caption{The pole terms $\langle v_1,v_2\rangle$ from the factor $\langle V_4,
    V_1\rangle$. There is a summation over $j$ from $r$ to $n_2-1$.}
    \label{figcg}
   \end{figure}

The last piece of the pole terms is from the vanishing of $\lan
V_4, V_1 \ran$ in Fig.~\ref{figcf} and Fig.~\ref{figcg} by setting
$V_1=v_1$ and $V_4=-v_1-v_2$ and summing over $j$. The
contribution corresponds to the Fig.~\ref{figcf} is
\begin{eqnarray}
C_4={(\a_1+\a_2)^3\over \a_1\a_2} \lan v_0,\sum_{j=n_1+1}^{r-1}
\vp_j \lambda_j-\vp_r \sum_{m=n_1+1}^{r-1}\lambda_m\ran ,
\end{eqnarray}
and the contribution corresponds to the Fig.~\ref{figcg} is
\begin{eqnarray}
C_5={(\a_1+\a_2)^3\over \a_1\a_2} \lan v_0, \vp_r
\sum_{m=n_1+1}^r\lambda_j+ \sum_{j=r+1}^{n_2} \vp_j \lambda_j\ran.
\end{eqnarray}

By summing all these five contributions together, we have
\begin{eqnarray}
C_1+C_2+C_3+C_4+C_5={(\a_1+\a_2)^3\over \a_1\a_2} \lan v_0,
\sum_{i=1}^n \vp_i\lambda_i \ran=0.
\end{eqnarray}

\begin{figure}[h!]
    \epsfxsize=100mm%
    \hfill\epsfbox{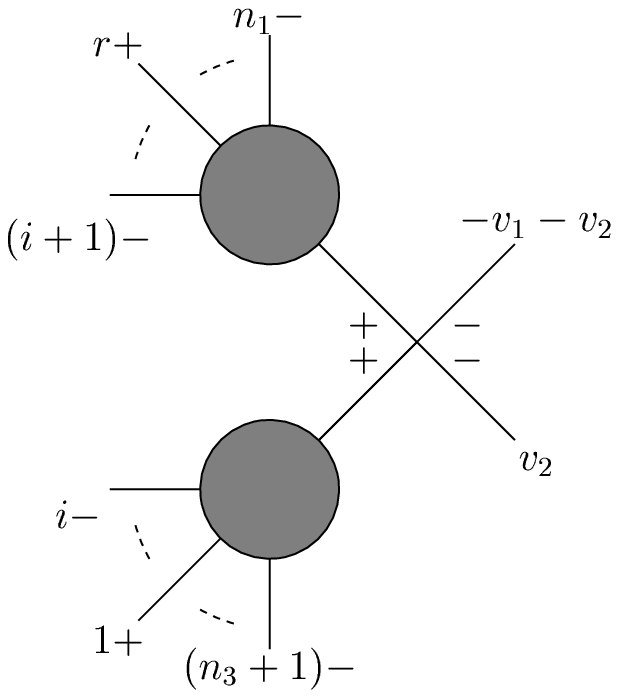}\hfill~\\
    \caption{All of the possible pole terms are come from this kind of diagrams
 when $3\le r\le n_1$, but actually there is no
    singularity.}
    \label{fignopole}
   \end{figure}

This prove that there is no pole terms for the function $F(\vp)$
when $n_1+1 \le r\le n_2$. The case for $n_2+1 \le r\le n_3$ is
similar. The case for $r=n$ or $r=2$ is also similar and the
algebra becomes  a little bit  easier because we need only
consider three different kinds of diagram in these two cases as
shown in \cite{Zhu}. When $3 \le r \le n_1$, the situation is even
easier because in this case all of the possible contribution to
the pole terms is come from the diagrams in fig.~\ref{fignopole},
but actually the numerator has a zero of order 4 while the
denominator has a zero of order 1. So there is no singularity in
every individual term. Similarly there is no pole terms when
$n_3+1\le r\le n-1$. In summary we proved that all the finite pole
terms in $F(\vp)$ are vanishing in all different cases.

So $F(\vp)$ is independent of $\vp_j$ for $2\le j  \le n$, $j \ne
r$. Now we compute $F(\vp)$ explicitly by choosing a special set
of $\vp_j$ for $2 \le j \le n$, $j \ne r$.\footnote{The case when
$r=n$ or $r=2$ is a little bit different, and it is done in
\cite{Zhu}.} A convenient choice is as follows:
\begin{equation}
\vp_2=\cdots=\vp_{r-1}=x,  \vp_{r+1}=\cdots=\vp_n=y.
\end{equation}
By using eq.~(\ref{eqla1}) and eq.~(\ref{eqla2}), we have
\begin{equation}
\lambda_1 = { (x-\varphi_r)\lambda + (y-\varphi_r)\mu \over
\varphi_r - \varphi_1}, \qquad \lambda_r =  - {
(x-\varphi_1)\lambda + (y-\varphi_1)\mu \over \varphi_r -
\varphi_1}, \label{eq38}
\end{equation}
by setting $\sum_{i=2}^{r-1}\lambda_i = \lambda$ and
$\sum_{j=r+1}^n \lambda_j= \mu$. In the following we will use the
new variables $x'$ and $y'$ which  we define as
\begin{equation}
x^{\prime}={x-\vp_r\over \vp_r-\vp_1}, \qquad
y^{\prime}={y-\vp_r\over \vp_r-\vp_1},
\end{equation}
then we have
\begin{equation} {x-\vp_1\over
\vp_r-\vp_1}=x^{\prime}+1, \qquad {y-\vp_1\over
\vp_r-\vp_1}=y^{\prime}+1 .
\end{equation}

Now  we show that for generic $x$ and $y$ (and generic $\lambda_i$
which we don't say explicitly) all the possible $\langle
V_1,V_2\rangle$, $\lan V_2,V_3\ran$, $\langle V_3,V_4\rangle$ and
$\langle V_4,V_1\rangle$ for different choices of $i$, $j$, $k$
and $l$ are non-vanishing.

In order to do  this, let's  define a matrix $\{t_{ij}\}_{(1\le
i\le n, 1\le j\le 4)}$. If $\lambda_i$ is one of the terms in
$V_j$ then we define $t_{ij}=1$. Otherwise we set $t_{ij}=0$. Then
 \begin{eqnarray}
 V_j & = & \sum_{i=1}^n
t_{ij}\lambda_i=\sum_{i=2}^{r-1}\lambda_i[t_{ij}+t_{1j}x^{\prime}-t_{rj}(x^{\prime}+1)]
\nonumber \\
& &
+\sum_{i=r+1}^{n}\lambda_i[t_{ij}+t_{1j}y^{\prime}-t_{rj}(y^{\prime}+1)].
\end{eqnarray}
So for generic $x$, $y$ and $\lambda_i$, $V_i$ and $V_{i+1}$ are
linearly independent and so  $\lan V_i, V_{i+1}\ran$ is
non-vanishing.

For our choice of $\varphi$, the possible non-vanishing factor for
$(\varphi_i - \varphi_{i+1}) (\varphi_j - \varphi_{j+1})(\varphi_k
- \varphi_{k+1})$ is from $i=1$, $j =r-1$ , $k=r$ and $l=n$ only.
Then we have
\begin{eqnarray}
 F(\vp) & = & (\vp_1-x)(x-\vp_r)(\vp_r-y)(y-\vp_1)\nonumber \\
& &  \times{\lan\lambda,
 \mu \ran\over \lan \lambda_1,\lambda\ran
\lan\lambda, \lambda_r\ran \lan \lambda_r, \mu \ran \lan \mu,
\lambda_1 \ran}.
\end{eqnarray}
By using  eq.~(\ref{eq38}), we have
\begin{eqnarray}
\lan \lambda_1, \lambda\ran&   = &-{y-\vp_r\over \vp_r-\vp_1}\lan
\lambda, \mu \ran, \\
  \lan \lambda, \lambda_r\ran&   = &-{y-\vp_1\over \vp_r-\vp_1}
\lan
\lambda, \mu \ran, \\
  \lan \lambda_r, \mu\ran&   = &-{x-\vp_1 \over \vp_r-\vp_1} \lan
\lambda, \mu \ran, \\
  \lan \mu, \lambda_1\ran&   = &-{x-\vp_r \over \vp_r-\vp_1} \lan
\lambda, \mu \ran ,
\end{eqnarray}
and so we have
\begin{equation}
F(\varphi) = (\varphi_r-\varphi_1)^4 .
\end{equation}
This completes the proof of eq.~(\ref{eqac}).

\section*{Appendix B: The computation of $P(\Gamma)$}

In this appendix, we will give  the proof of eq.~(\ref{eqgamma}).
We assume that the off-shell graviton has positive helicity. The
case when it has negative helicity is similar.

 \begin{figure}[ht]
    \epsfxsize=100mm%
    \hfill\epsfbox{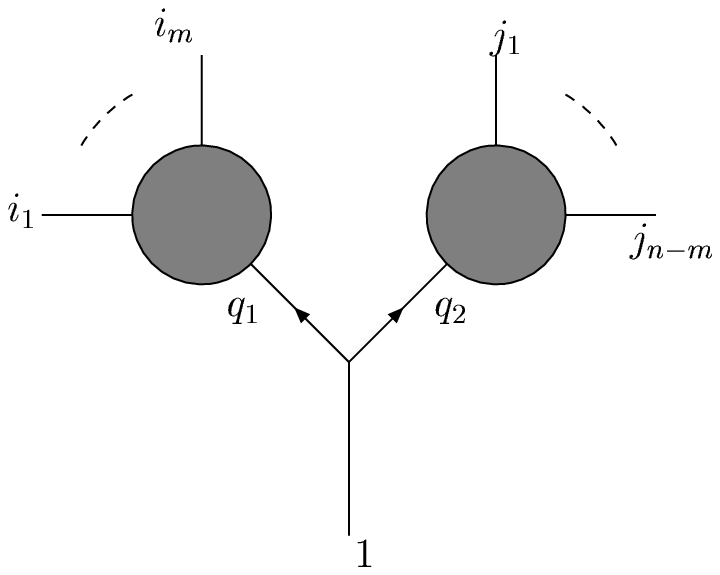}\hfill~\\
    \caption{The decomposition for the amplitude for gravitons using in
    the computation of
    $P(\Gamma)$.
    }
    \label{figmultigraviton}
   \end{figure}

When $n=3$, we have $\lambda_1+\lambda_2\phi_2+\lambda_3
\phi_3=0$, so $\langle 1,2 \rangle=\phi_3\langle 2,3 \rangle$,
$\langle 3,1 \rangle=\phi_2 \langle 2,3 \rangle$. Then
\begin{equation}
A_3(1+,2-,3-)={1 \over \phi_2^2 \phi_3^2}\langle
2,3\rangle^2={p_1^2 \over \phi_2^2 \phi_3^2}{\langle 2,3\rangle
\over [2,3]}.
\end{equation}
There is only one inequivalent undirected tree with vertex set
$\{2,3\}$. There is only one edge $e_{23}$ in this tree. It is
easy to see that
\begin{equation}
P(\Gamma)=p(2)p(3)p(e_{23})=\phi_2^{-4}\phi_3^{-4} \left( \phi_2^2
\phi_3^2 {\langle 2,3\rangle \over [2,3]}\right)={1 \over \phi_2^2
\phi_3^2}{\langle 2,3\rangle \over [2,3]}.
\end{equation}
So $M_3(1+,2-,3-)=p_1^2 P(\Gamma)$. This completes the proof for
eq.~(\ref{eqgamma}) when $n=3$. Assuming that it is true for all
$k\le n$, we will prove that it is also true for $k=n+1$.

We use the diagram decomposition in Fig.~\ref{figmultigraviton} to
calculate $A_{n+1}$. So we have
\begin{eqnarray}
A_{n+1}& = &\sum_{\{i_1,\cdots,
i_m\}}A_{m+1}(-q_1,i_1,\cdots,i_m)\, {1 \over q_1^2} \nonumber \\
& & \times A_{n-m+1}(-q_2,j_1,\cdots,j_{n-m}) \, {1\over q_2^2} \,
V_3(1,q_1,q_2),
\end{eqnarray}
where $i_1,\cdots,i_m$ are any $m\,(1\le m\le n-1)$ gravitons in
$\{2,\cdots,n+1\}$, $j_1,\cdots,j_{n-m}$ are the remaining $n-m$
ones. $q_1$ and $q_2$ are two internal lines with momenta
$q_1=\sum_{k=1}^mp_{i_k}$ and $q_1=\sum_{k=1}^{n-m}p_{j_k}$
respectively. We denote  $\{i_1,\cdots,i_m\}$ as $S_1$ and
$\{j_1,\cdots,j_{n-m}\}$ as $S_2$. It is easy to see that when
$m=1$ ($m=n-1$), if we understand $A_{2}(-q_1,i_1)$
($A_{2}(-q_2,j_1)$) as $1/\phi_{i_1}^4$ ($1/\phi_{j_1}^4$), then
these two degenerate cases are included correctly.

By using the assumed result for all less multi-graviton
amplitudes, we can get
\begin{eqnarray}
A_{n+1}=\sum_{\Gamma_1,\Gamma_2} P(\Gamma_1) P(\Gamma_2)
V_3(1,q_1,q_2),
\end{eqnarray}
where $\Gamma_1$ and $\Gamma_2$ are two disjoint undirected trees
subject to the constraint that the union of their vertex set are
$\{2,3,\cdots,n\}$. (In this appendix, the summation over
$\Gamma_1$ and $\Gamma_2$ is always understood to be under this
constraint.) When $m=1$ ($m=n-1$) the tree $\Gamma_1$ ($\Gamma_2$)
includes only one vertex and the there are no edges in $\Gamma_1$
($\Gamma_2$). These degenerate cases are included correctly, too.

From eq.~(\ref{eq3gravitons}), we have
\begin{eqnarray}
V_3(1,q_1,q_2)&=&\langle
\lambda_{q_1},\lambda_{q_2}\rangle^2=\left(\sum_{i\in
V_1,j\in V_2}\langle \lambda_{p_i},\lambda_{p_j} \rangle\right)^2\nonumber\\
&=& \sum_{i,j\in V_1, k,l\in V_2} \langle
\lambda_{p_i},\lambda_{p_k} \rangle \langle
\lambda_{p_j},\lambda_{p_l}\rangle   ,\label{eqsquart}
\end{eqnarray}
where $V_i$ is the vertex set $V(\Gamma_i)$ of graph $\Gamma_i$
($i=1,2$). So we have
\begin{equation}
A_{n+1}=\sum_{\Gamma_1,\Gamma_2}
P(\Gamma_1)P(\Gamma_2)\sum_{i,j\in V_1, k,l\in V_2} \langle
\lambda_{p_i},\lambda_{p_k} \rangle \langle
\lambda_{p_j},\lambda_{p_l}\rangle.
\end{equation}
By denoting $2 p_i\cdot p_j=\langle i,j\rangle [i,j]$ as $s_{ij}$
and then using
\begin{equation}p_1^2=\left( \sum_{i=2}^{n+1}p_i \right)^2=\sum_{2\le
i<j\le n+1}s_{ij},
\end{equation}
we   get
\begin{equation}
p_1^2\sum_{\Gamma,V(\Gamma)=S(n+1)}P(\Gamma)=\sum_{\Gamma,V(\Gamma)=S(n+1)}P(\Gamma)
\sum_{i,j\in V(\Gamma), i<j}s_{ij} .
\end{equation}

For any $i,j\in V(\Gamma)$, there is an unique path in the tree
$\Gamma$ jointing $i$ and $j$. We denote this path as
$v_1=i,v_2,\cdots,v_{k-1},v_k=j$. By using the following identity:
\begin{equation}
{[i,j]\over \phi_i\phi_j}={[i,k]\over \phi_i\phi_k}+{[k,j]\over
\phi_k\phi_j},
\end{equation}
we get
\begin{equation}
{[i,j]\over \phi_i\phi_j}=\sum_{s=1}^{k-1}{[s,s+1]\over
\phi_{v_s}\phi_{v_{s+1}}}.
\end{equation}
So \begin{eqnarray} P(\Gamma)s_{ij}&=&P(\Gamma)\langle i,j\rangle
\phi_i\phi_j {[i,j]\over \phi_i \phi_j}\nonumber\\
&=&\sum_{s=1}^{k-1}P(\Gamma)\langle i,j\rangle
\phi_i\phi_j{[v_s,v_{s+1}]\over \phi_{v_s}\phi_{v_{s+1}}} .
\label{eqgamma2}
\end{eqnarray}

If we move away an edge from a tree, we can get two disjoint
trees. Conversely, assuming that $\Gamma_1$ and $\Gamma_2$ are two
disjoint trees, if we  connect one vertex in $\Gamma_1$ with
another vertex in $\Gamma_2$ by an edge, we get a bigger  tree. We
denote the two trees obtained by moving away the edge $e_{v_s
v_{s+1}}$ as $\Gamma_1(v_s)$ and $\Gamma_2(v_s)$. Then we have
\begin{eqnarray} P(\Gamma)s_{ij}
&=&\sum_{s=1}^{k-1}P(\Gamma_1(v_s)) P(\Gamma_2(v_s)){\langle
v_s,v_{s+1}\rangle \over
[v_s,v_{s+1}]}\phi_{v_s}^2\phi_{v_{s+1}}^2\langle i,j\rangle
\phi_i\phi_j{[v_s,v_{s+1}]\over \phi_{v_s}\phi_{v_{s+1}}}\nonumber \\
&=&\sum_{s=1}^{k-1}P(\Gamma_1(v_s)) P(\Gamma_2(v_s)){\langle
v_s,v_{s+1}\rangle }\phi_{v_s}\phi_{v_{s+1}}\langle i,j\rangle
\phi_i\phi_j.
\end{eqnarray}
From this we can get
\begin{eqnarray}
\sum_{\Gamma,V(\Gamma)=S(n+1)}P(\Gamma)p_1^2&=&\sum_{\Gamma,V(\Gamma)=S(n+1)}
P(\Gamma)\sum_{i,j \in V(\Gamma), i<j}
s_{ij}\nonumber\\&=&\sum_{\Gamma_1,\Gamma_2}
P(\Gamma_1)P(\Gamma_2)  \sum_{i,j\in V_1, k,l\in V_2} \langle
\lambda_{p_i},\lambda_{p_k} \rangle
\langle \lambda_{p_j},\lambda_{p_l}\rangle\nonumber\\
&=&A_{n+1},
\end{eqnarray}
as announced. This completes the proof of eq.~(\ref{eqgamma}) by
mathematical induction.

\end{document}